\documentclass[reprint]{revtex4-2}

\usepackage{xcolor}
\usepackage[british]{babel}
\usepackage[utf8]{inputenc} 
\usepackage{epsfig} 
\usepackage{graphicx}
\usepackage{xcolor}
\usepackage{physics}
\usepackage{enumitem}

\usepackage{subfigure}
\usepackage{tikz}
\usetikzlibrary{shapes,arrows,arrows.meta,positioning,calc}
\usepackage{pgf}
\usepackage{ifpdf}

\ifpdf
  \usetikzlibrary{arrows.meta}
\else
  \PassOptionsToPackage{dvips}{graphicx} 
\fi
\tikzset{
    line/.style = {
        draw,
        -{Latex[round,length=20pt,width=6pt]} 
    },
    cloud/.style = {
        draw, ellipse, node distance=2.5cm,
        minimum height=2em
    }
}

\usepackage{amsmath,amssymb,amsthm,mathtools}\usepackage{bm}
\usepackage{geometry}
\usepackage{microtype}
\usepackage[colorlinks=true,linkcolor=blue,citecolor=blue,urlcolor=blue]{hyperref}
\theoremstyle{remark}

\begin{document}

\title{Can Quantum Field Theory be Recovered from Time-Symmetric Stochastic Mechanics? Part I: Generalizing the Liouville Equation}
\author{Simon Friederich, Mritunjay Tyagi}
\email{s.m.friederich@rug.nl, m.tyagi@rug.nl}
\affiliation{University of Groningen, University College Groningen, Hoendiepskade 23/24, 9718BG Groningen, the Netherlands}

\begin{abstract}
We explore whether quantum field theory can be understood as the statistical mechanics of a time-reversal-invariant stochastic generalization of Hamiltonian dynamics. The motivation for this project, started with this paper, is to assign sharp values to all observables and thereby avoid the quantum measurement problem. In classical mechanics, motion is deterministic and corresponds to an evolution of the phase space probability density according to Liouville's equation that is governed by first derivatives of the Hamiltonian in phase space. We derive a generalization of the Liouville equation with natural constraints -- namely, reduction to classical Hamiltonian dynamics as the stochasticity parameter $\hbar\mapsto0$, Fokker-Planck form for the probability density evolution, local Hamiltonian dependence, time-reversal invariance, energy conservation, and minimality -- which turns out to be a Fokker-Planck equation with a generalized diffusion matrix that is symmetric, traceless, and constructed from the Hessian of the Hamiltonian. We then show that the Schr\"odinger equation in the coherent-state phase-space formulation of certain bosonic QFTs has precisely this form, with the Husimi function playing the role of the phase space probability density. The question to what extent this equation can be interpreted in terms of objective stochastic field theories is discussed in a companion paper.
\end{abstract}

\maketitle

\section{Introduction}\label{sec:intro}
The quest for a compelling understanding of quantum theory as conceptually similar to classical statistical mechanics is almost as old as quantum theory itself. In such an understanding, quantum states would be akin to probability distributions on phase space, and all dynamical variables would have sharp values at all times. Einstein famously expected that future research would reveal this to be the case: ``I am rather firmly convinced that the development of physics will be of this type; but the path will be lengthy and difficult.'' (\cite{einstein1949}, p.\ 672)

Since Einstein wrote these words, a large variety of no-go theorems such as Bell's theorem \citep{bell1964,myrvold2024,goldstein2011}, the Kochen-Specker theorem \cite{kochen1967,held2022}, and the PBR theorem \cite{pusey2012,leifer2014}, have discouraged many regarding the prospects for realizing Einstein's hope and vision. These no-go results constrain trajectory interpretations within specific frameworks -- particularly the ontological models framework that assumes Markovian dynamics. As we show in a companion paper, the stochastic dynamics that emerge from the evolution equation derived here turn out to be fundamentally non-Markovian when analyzed at the micro-level, which places them outside the scope of these theorems. However, whether this formal evasion translates into a viable physical interpretation remains an open and subtle question that we address separately.

Here we investigate a different, more positive route rather than focusing on no-go results as our starting point: start with Newtonian mechanics and generalize it in a controlled way, subject to certain well-motivated constraints, and see whether any realistic quantum theories -- ideally the most fundamental quantum field theories for which we have robust empirical evidence -- correspond to the resulting framework.

Unlike classical mechanics, quantum theory, via the Born rule, makes inherently statistical predictions. When seeking generalizations of classical dynamics that reproduce quantum behaviour it is therefore natural to implement a statistical element by allowing the dynamics to be stochastic. However, as we will see in Section~\ref{sec:challenge}, adding stochastic terms to the classical Hamilton's equations in what may appear to be the most straightforward way, which corresponds to including a diffusion term in Liouville's equation, introduces a radical time asymmetry that has no correspondence in the Schr\"odinger dynamics of quantum theory. Nelson's stochastic mechanics, formulated in configuration space, avoids this problem by making the diffusion term dependent on the wave function. But this move is incompatible with the Einsteinian vision pursued here, where the (modulus squared of the) wave function is understood as conceptually equivalent to the probability density in classical statistical mechanics, so we will seek a different approach.

In Section~\ref{sec:derivation} we present a novel solution to this challenge by imposing various physically motivated constraints on the time evolution equation for the probability density $\rho(\mathbf z,t)$ in phase space for our desired stochastic generalization of classical mechanics. The first constraint (C1) is that the Liouville equation of classical statistical mechanics, based on deterministic Hamiltonian motion, should be recovered in the limit where some parameter $\hbar$ which intuitively measures ``stochasticity'', i.e. deviation from determinism, goes to zero. Our second constraint, motivated by the idea that trajectories should be continuous, is that the time evolution of the probability density should be described by a Fokker-Planck equation, which can be seen as the minimal extension of deterministic dynamics in the stochastic case. Our third constraint is that all parameters in the evolution equation should be built completely from the local Hamiltonian. This rules out, in particular, that (features of) the probability density $\rho(\mathbf z,t)$ can itself become a dynamical agent, as in, for example, Bohmian mechanics or Nelsonian mechanics.

Our fourth constraint (C4) is time reversal invariance. This turns out to imply that the diffusion matrix in the Fokker-Planck equation cannot be an ordinary positive semi-definite diffusion matrix but that it must rather be indefinite, with zero trace. Our fifth constraint (C5), energy conservation on expectation, further narrows down the form of the stochastic contribution: Namely, its generalized diffusion matrix must be Frobenius orthogonal to the Hessian of the Hamiltonian. Our sixth constraint (C6), minimality, dictates the precise form of the generalized diffusion matrix in terms of entries of the Hessian, up to a global constant. We present the resulting time evolution equation for the probability density in both complex and real phase space coordinates. Finally, we propose a seventh constraint (C7) that fixes the length scales $\ell_j$ used in the definition of the complex phase space coordinates by aligning the two dichotomies free/interacting and deterministic/stochastic.

Next, in Section~\ref{sec:Drummond}, we show that, for a significant class of bosonic quantum field theories, the time evolution equation for the probability density derived in Section~\ref{sec:derivation} agrees exactly with the Schr\"odinger equation for the quantum state in the coherent state representation (Husimi function $Q$). The fact that the evolution of the Husimi function in those theories follows a Fokker-Planck equation with a traceless diffusion matrix was noted earlier by Drummond \citep{drummond2021}. Our result clarifies how the parameters in this equation relate to the gradient and the Hessian of the Anti-Wick symbols of the Hamiltonian operator. Drummond suggests that an interpretation of that evolution equation in terms of stochastic field trajectories, which would solve the quantum measurement problem, is available. We critically examine this claim and the dynamics of stochastic field trajectories in a companion paper.

Finally, in Section~\ref{sec:standard_model} we consider the limitations of the preceding analysis. The evolution equation for the Husimi function has a Fokker-Planck form in free theories and in interacting theories with a quartic density-density coupling in the complex phase space variables. We show that the Bose-Hubbard model is included  among these theories, whereas real and complex scalar field theories and the gauge boson self-interactions in non-Abelian gauge theories are not. Including arbitrary interaction terms would require going beyond the Fokker-Planck ansatz. We consider possible reactions to this limitation of the present analysis: that it points to fundamental limitations in the project of understanding quantum field theories as the statistical mechanics of stochastic theories; that one can extend the Fokker-Planck equation-based approach by adding phase space degrees of freedom and replace coherent state projectors by (more general) Gaussian bosonic operators \citep{corneydrummond}; and that one proposes the theories with evolution equation of the Husimi function truncated at Fokker-Planck level as empirical alternative rival theories to the theories in the Standard Model of elementary particle physics.

The paper concludes in Section \ref{sec:summary} with a summary and an outlook on further steps and challenges.

\section{The challenge: stochasticity versus time-reversal invariance}\label{sec:challenge}

As outlined in the introduction, our goal is to recover quantum theory -- preferably at least the most reliable quantum field theories for which we have strong empirical evidence -- as the statistical mechanics of stochastic generalizations of classical Hamiltonian mechanics. In this section, we explain why the naive addition of stochastic terms introduces a radical time asymmetry that is incompatible with the ``time democracy'' of the Schr\"odinger equation in quantum theory. This will motivate us in the following section, Section~\ref{sec:derivation}, to explicitly impose time reversal invariance of the evolution equation for the phase space probability density as one of our key constraints.

\subsection{Watanabe's general argument}

The incompatibility between standard stochastic dynamics and time-reversal invariance can be seen straightforwardly based on a simple yet powerful argument by Watanabe \citep{watanabe1965}. Consider any stochastic theory in which the dynamics are described by a forward-time Markov kernel $P(x',t+\Delta t \mid x,t)$, where $x$ represents the system's state in some suitable space. Suppose this kernel depends only on the parameters of the theory (such as those that appear in its Hamiltonian) and not on the probability distribution $P(x,t)$ itself, i.e. it is fixed as soon as those parameters are fixed.

By Bayes' theorem, the time-reversed conditional probability is given by
\begin{equation}
P(x,t | x',t+\Delta t) = \frac{P(x',t+\Delta t | x,t)\, P(x,t) }{ P(x',t+\Delta t)}\,.\label{Bayes}
\end{equation}

The crucial observation is this: if the forward conditional probability $P(x',t+\Delta t | x,t)$ is fixed purely by the theory's parameters (independent of how the system was prepared), while the unconditional probability $P(x,t)$ \emph{does} depend on the preparation procedure, then the backward conditional probability $P(x,t | x',t+\Delta t)$ must also depend on the preparation. This follows immediately from Eq.\ (\ref{Bayes}).

The forward and backward conditional probabilities thus have fundamentally different character: one is purely dynamical and ``law-like'', the other is preparation-dependent. Unless the dynamics is deterministic (all probabilities 0 or 1), this asymmetry is unavoidable. Any stochastic theory that makes the conditional probabilities in \emph{one} direction law-like (fixed by fixing the theory's parameters), while the unconditional probabilities remain preparation-dependent, must therefore treat the two time directions in radically different ways and, in particular, cannot be time-reversal invariant. This leaves three options: determinism (where all conditional probabilities are trivially law-like), making the unconditional probabilities themselves law-like (as in Nelson's approach, see below), or -- the route pursued in the Part II companion paper to the present one -- treating \emph{neither} forward- nor backward-oriented conditional probabilities as fully law-like.

\subsection{Illustration: the Fokker-Planck equation}

This general tension between stochasticity and time reversal invariance can be nicely illustrated by the standard framework for continuous stochastic dynamics. Consider a system whose state $\mathbf{x}$ evolves according to an It\^{o} stochastic differential equation (SDE):
\begin{equation}
d x_i(t)= a_i(\mathbf{x})\,dt + c_{ij}(\mathbf{x}) d W_j\,, \label{langevin}
\end{equation}
where $a_i(\mathbf{x})$ is the drift, $c_{ij}(\mathbf{x})$ describes the diffusion, and $dW_j$ are the Gaussian-distributed increments of independent Wiener processes. At the level of the probability density $\rho(\mathbf{x},t)$, this corresponds to a Fokker-Planck equation:
\begin{eqnarray}
\frac{\partial \rho(\mathbf{x},t)}{\partial t}&=&-\sum_i\frac{\partial }{\partial x_i}\bigl[a_i(\mathbf{x},t) \rho(\mathbf{x},t)\bigr] \nonumber\\&+& \sum_{i,j}\frac{1}{2} \frac{\partial^2}{\partial x_i\partial x_j}\bigl[D_{ij}(\mathbf{x},t) \rho(\mathbf{x},t)\bigr]\,,
\label{Fokker_Planck}
\end{eqnarray}
where the diffusion matrix $D_{ij}$ satisfies
\begin{equation}
D_{ij} = \sum_{k} c_{ik}c_{jk}\,.
\end{equation}
For Eq.\ (\ref{langevin}) to define a well-posed continuous Markov process, the symmetric matrix $D_{ij}$ must be positive semi-definite.

This positive-definiteness of the diffusion matrix encodes the time asymmetry of the dynamics. If a sharply localized initial condition $\rho(\mathbf{x},t_i) = \delta(\mathbf{x}-\mathbf{x}_0)$ is specified, Eq.\ (\ref{Fokker_Planck}) yields a well-posed forward evolution. However, specifying a sharp final condition leads to an ill-posed problem \citep{miranker1961}. The probability distribution ``spreads'' forward in time but not backward.

For small $\Delta t$, the forward conditional probability takes the approximate form
\begin{eqnarray}
\rho(\mathbf{x}', t+\Delta t \mid \mathbf{x}, t)
&=& \frac{1}{\sqrt{(2\pi \Delta t)^{N} \det D}}
\nonumber\\&&\exp\!\left[
-\frac{|\Delta \mathbf{x} - \mathbf{a}\Delta t|^2_{D^{-1}}}{2\Delta t}
\right]\,,
\end{eqnarray}
where $\Delta \mathbf{x} = \mathbf{x}' - \mathbf{x}$ and $|\mathbf{v}|^2_{D^{-1}} \equiv \mathbf{v}^{\!\top} D^{-1} \mathbf{v}$ denotes the squared norm with metric $D^{-1}$ and $N$ is the dimensionality of $\mathbf{x}$. If both drift $\mathbf{a}$ and diffusion $D$ depend only on the system's Hamiltonian or other dynamical parameters, this forward probability is purely dynamical, i.e. it does not depend on the initial condition $\rho(\mathbf x_i,t)$. Applied to this situation, Watanabe's argument (Eq.\ \ref{Bayes}) makes it clear that the backward probability $\rho(\mathbf{x},t | \mathbf{x}',t+\Delta t)$ must depend on the preparation-dependent $\rho(\mathbf{x},t)$ -- manifesting the time-reversal asymmetry.

\subsection{The Nelson-F\'enyes approach and its limitations}

It is instructive to examine how the Nelson-F\'enyes stochastic mechanics \citep{fenyes1952,nelson1966,bacciagaluppi2012}, an approach to quantum foundations which can be seen as a stochastic version  of Bohmian mechanics, achieves time-reversal invariance. This approach works in configuration space and makes the wave function $\psi$ itself a dynamical agent. Writing $\psi(x,t)=\sqrt{R(x,t)}\,e^{iS(x,t)/\hbar}$, the forward conditional probability becomes
\begin{multline}
P(x',t+\Delta t\,|\,x,t)\;\sim\;
\bigl(2\pi \tfrac{\hbar}{m}\Delta t\bigr)^{-N/2}\\
\times\exp\!\biggl[
-\frac{1}{2\,\tfrac{\hbar}{m}\Delta t}
\Bigl(x'-x\\
-\Delta t\,\bigl[\tfrac{1}{m}\nabla S(x,t)+\tfrac{\hbar}{2m}\nabla\ln R(x,t)\bigr]\Bigr)^{\!2}
\biggr]\,,
\end{multline}
where $R$ and $S$ evolve according to the Madelung form of the Schr\"odinger equation. Here the drift explicitly depends on the parameters $R$ and $S$ that appear in the wave function $\psi$. In effect, since $R$ and $S$ are objective physical quantities, since $R^2$ also plays the role of the probability density, that probability is directly identified with an objective physical quantity, so it has itself been made ``law-like.'' This move enables side-stepping Watanabe's argument, treat both forward- and backward-oriented conditional probabilities as law-like and, thereby, allows one to restore time-reversal symmetry. (See \citep{grabert1979} for a critique of Nelsonian mechanics that points out the disanalogy with a genuine diffusion problem.)

However, this resolution is incompatible with the Einsteinian vision we pursue. In classical statistical mechanics, probability distributions can be thought of as corresponding to ensembles, not individual systems. Alternatively one can see them as epistemic tools that reflect our incomplete knowledge of microscopic states. Giving $\rho$ (or equivalently $\psi$) a \emph{dynamical} role -- making its gradients appear in the equations of motion for an individual system -- represents a radical conceptual departure. The probability density ceases to be merely a description of our knowledge and becomes instead a physical field influencing particle motion. From the point of view of classical statistical mechanics this can be seen as a ``category mistake.''

Moreover, the fact that the wave function $\psi(x,t)$ is a field on the $3N$-dimensional configuration space introduces another problem. A subsystem's motion becomes instantaneously sensitive to arbitrarily distant regions through this global ``pilot wave'' field. This nonlocality differs fundamentally from the position-dependent forces of Newtonian mechanics, where interactions depend only on the actual positions of distant bodies, not on a configuration-space field with autonomous dynamics. This feature also poses severe obstacles for relativistic generalization \citep{wallstrom1994}.

\subsection{The path forward}

As we have seen, achieving time-reversal invariance in a stochastic generalization of classical mechanics presents a non-trivial challenge. We propose that a solution should be sought that combines the following three key requirements:
\begin{enumerate}
\item \textbf{Continuous trajectories}: The evolution equation should have Fokker-Planck form, which is the natural framework for continuous stochastic dynamics.
\item \textbf{Time-reversal invariance}: A natural requirement to match the structure of quantum dynamics. 
\item \textbf{Dynamical parameters depend only on the Hamiltonian}: The probability density $\rho$ should remain purely epistemic, not become a dynamical agent. This suggests that dynamical parameters should be functions of the Hamiltonian. (Whether the resulting evolution equation permits a consistent trajectory interpretation is addressed in our companion paper.)
\end{enumerate}

Watanabe's argument makes it clear that these requirements are incompatible with Markovian approaches where the forward conditional probabilities are fixed by the dynamical parameters while the unconditional probability density depends on the chosen preparation. We now turn to the systematic derivation of an alternative framework, which matches our requirements, by successively imposing plausible physical constraints.

\section{Deriving the time-symmetric evolution of $\rho$}\label{sec:derivation}

In this section, we derive our time-symmetric stochastic generalization of classical mechanics subject to physical constraints. Statistical mechanics is conveniently formulated in phase space, so we take the Hamiltonian, phase space, formulation of classical mechanics as our starting point. 

We work in phase space with complex phase space coordinates
\begin{equation}
    \alpha_i = \sqrt{\frac{\kappa}{2}}\,\left(\frac{q_i}{\ell_i}+ i\frac{\ell_i\,p_i}{\kappa}\right)\,,\label{complex_variable}
\end{equation}
where $i=1,\ldots, N$ labels degrees of freedom, $\kappa$ is an action scale (which we will later identify with $\hbar$), and the $\ell_i$ are length scales that can differ by degree of freedom. The index $i$ may label momentum modes, lattice sites, internal degrees of freedom (such as colour), or combinations thereof. For a complex field, each spatial or momentum mode contributes two sets of phase-space variables corresponding to particle and antiparticle excitations. The length scales $\ell_i$ are, at this stage, free parameters of the theory; we will return to their determination in Section~\ref{subsec:scale_fixing}. Throughout this paper, we work with a finite number of degrees of freedom, corresponding to a lattice regularization of the field theory with a fixed spatial cutoff. The extension to continuous field theories on non-compact spaces raises additional mathematical difficulties -- notably, the phase space is no longer locally compact, and the existence and uniqueness of classical evolution equations in that setting is itself non-trivial.

We use the compact notation
\begin{equation}
    \mathbf{z}=\begin{bmatrix}
\boldsymbol{\alpha} \\
\boldsymbol{\alpha}^* 
\end{bmatrix}\label{phase_space_vector}
\end{equation}
for the full $2N$-dimensional phase-space vector.

Adding stochastic terms to Hamilton's equations means turning them into stochastic differential equations (SDEs). However, any given stochastic process can be expressed by various equivalent SDEs -- for instance, continuous Markov processes can be expressed equivalently in It\^{o} or Stratonovich form. To avoid this non-uniqueness, we focus instead on the time evolution equation for the \emph{probability density} $\rho(\mathbf{z},t)$. We take the Liouville equation as our starting point and impose our constraints successively to determine the unique form of the stochastic generalization.

\subsection{Classical limit and Fokker-Planck form}
Our first constraint is:
\begin{itemize}
\item \textbf{(C1) Classical limit}: The evolution equation for $\rho(\mathbf{z},t)$ consists of a Liouville term plus a stochasticity term. The latter vanishes as the stochasticity parameter $\hbar \to 0$.
\end{itemize}

This constraint ensures that, as $\hbar\to 0$, the time evolution reduces to the Liouville equation for classical statistical mechanics:
\begin{multline}
\frac{\partial \rho(\mathbf{z},t)}{\partial t}
= - \nabla_{\mathbf{z}}\!\cdot\!\bigl(\mathbf{a}(\mathbf{z})\,\rho(\mathbf{z},t)\bigr)\\
= i\sum_i\!\left[
\frac{\partial}{\partial \alpha_i}\!\left(\frac{\partial H}{\partial \alpha_i^*}\rho\right)
- \frac{\partial}{\partial \alpha_i^*}\!\left(\frac{\partial H}{\partial \alpha_i}\rho\right)
\right]\!,\label{Liouville}
\end{multline}
where $\mathbf{a}(\mathbf{z})$ is the drift vector determined by Hamilton's equations.

Our second constraint is inspired by the hope that trajectories should remain continuous when stochasticity is introduced. For standard Markov processes, continuity is equivalent to Fokker-Planck evolution -- i.e., evolution equations involving at most second derivatives in the phase-space coordinates. However, as established in Section~\ref{sec:challenge}, standard Markov processes with positive-definite diffusion matrices are incompatible with time-reversal invariance. The resolution is to retain the Fokker-Planck form while allowing the diffusion matrix to be indefinite (having both positive and negative eigenvalues). In our companion paper, we discuss to what extent this structure can support continuous stochastic trajectories despite the abandonment of the standard Markov property. 

Our second constraint is thus:
\begin{itemize}
\item \textbf{(C2) Fokker-Planck form}: The evolution equation for $\rho(\mathbf{z},t)$ takes the form of a Fokker-Planck equation with a symmetric diffusion matrix $D_{ij}$, making no assumptions about the signs of its eigenvalues.
\end{itemize}

Together, constraints C1 and C2 determine the general structure:
\begin{widetext}
\begin{align}
\frac{\partial \rho(\boldsymbol{\alpha},\boldsymbol{\alpha}^*,t)}{\partial t}
&= i\sum_{i}
\left[
\frac{\partial}{\partial \alpha_i}
\Bigl(\frac{\partial H}{\partial \alpha_i^*}\,\rho\Bigr)
- \frac{\partial}{\partial \alpha_i^*}
\Bigl(\frac{\partial H}{\partial \alpha_i}\,\rho\Bigr)
\right]
\label{general_Fokker_Planck}\\[4pt]
&\quad
+ \frac{i\hbar}{2} \sum_{i,j}
\bigg\{
\frac{\partial^2}{\partial \alpha_i \partial \alpha_j}
\bigl(D_{ij}^{\alpha\alpha}\,\rho\bigr)
+ \frac{\partial^2}{\partial \alpha_i^* \partial \alpha_j^*}
\bigl(D_{ij}^{\alpha^*\alpha^*}\,\rho\bigr)
+ \frac{\partial^2}{\partial \alpha_i \partial \alpha_j^*}
\bigl((D_{ij}^{\alpha\alpha^*}+D_{ji}^{\alpha^*\alpha})\,\rho\bigr)
\bigg\}\!.\nonumber
\end{align}
\end{widetext}
Here the diffusion matrix has the block structure
\begin{equation}
D =
\begin{pmatrix}
D^{\alpha\alpha} & D^{\alpha\alpha^*} \\[4pt]
D^{\alpha^*\alpha} & D^{\alpha^*\alpha^*}
\end{pmatrix}\!,\label{diffusion_block_structure}
\end{equation}
with each block being an $N \times N$ matrix. We will occasionally decompose the right-hand side of Eq.\ (\ref{general_Fokker_Planck}) as $\mathcal{L}_L\rho + \mathcal{D}\rho$, where $\mathcal{L}_L\rho$ and $\mathcal{D}\rho$ denote the Liouville and diffusion contributions (first and second lines), respectively.

The diffusion matrix must be symmetric when expressed in real coordinates, which imposes constraints on its complex representation. Symmetry requires:
\begin{align}
D_{ij}^{\alpha\alpha}&=D_{ji}^{\alpha\alpha}\,,\label{symm1}\\
D_{ij}^{\alpha^*\alpha^*}&=D_{ji}^{\alpha^*\alpha^*}\,,\label{symm2}\\
D_{ij}^{\alpha\alpha^*}&=D_{ji}^{\alpha^*\alpha}\,.\label{symm3}
\end{align}

Additionally, since $\rho$ must remain real at all times and we have placed a factor $i$ in front of the diffusion term in Eq.\ (\ref{general_Fokker_Planck}), reality of the evolution requires:
\begin{align}
\left(D_{ij}^{\alpha\alpha}\right)^*&=-D_{ij}^{\alpha^*\alpha^*}\,,\label{reality1}\\
\left(D_{ij}^{\alpha\alpha^*}\right)^*&=-D_{ij}^{\alpha^*\alpha}\,.\label{reality2}
\end{align}

Combining Eqs.\ (\ref{symm3}) and (\ref{reality2}), we find that $D^{\alpha\alpha^*}$ is skew-Hermitian:
\begin{equation}
\left(D_{ij}^{\alpha\alpha^*}\right)^*=-D_{ji}^{\alpha\alpha^*}\,.\label{skew_hermitian}
\end{equation}

Using these identities, the diffusion contribution in Eq.\ (\ref{general_Fokker_Planck}) simplifies to
\begin{widetext}
\begin{equation}
\mathcal{D}\rho = \frac{i\hbar}{2} \sum_{i,j}
\left\{
\frac{\partial^2}{\partial \alpha_i \partial \alpha_j}
\bigl(D_{ij}^{\alpha\alpha}\,\rho\bigr)
- \frac{\partial^2}{\partial \alpha_i^* \partial \alpha_j^*}
\bigl((D_{ij}^{\alpha\alpha})^*\,\rho\bigr)
+ 2\frac{\partial^2}{\partial \alpha_i \partial \alpha_j^*}
\bigl(D_{ij}^{\alpha\alpha^*}\,\rho\bigr)
\right\}\!.\label{diffusion_simplified}
\end{equation}
\end{widetext}

In the following subsections, we impose additional physical constraints to determine the explicit form of the diffusion matrix $D_{ij}$.

\subsection{Local Hamiltonian dependence and time-reversal invariance}

Having established the general Fokker-Planck structure, we now impose constraints that determine the explicit form of the diffusion matrix. In the Liouville equation Eq.\ (\ref{Liouville}), the drift vector $\mathbf{a}(\mathbf{z})$ depends on the phase-space location $\mathbf{z}$ only through the Hamiltonian -- specifically, through its gradient. Notably, unlike in the F\'enyes-Nelson approach, the drift does not depend on the probability density $\rho$ itself. We preserve this feature by requiring:

\begin{itemize}
\item \textbf{(C3) Local Hamiltonian dependence}: All parameters in the evolution equation -- including both drift and diffusion terms --are constructed from the local Hamiltonian $H(\boldsymbol{\alpha},\boldsymbol{\alpha}^*)$ and its derivatives.
\end{itemize}

This constraint ensures that $\rho$ remains purely epistemic, describing our knowledge rather than influencing the dynamics. It applies in particular to the elements $D^{\alpha^{(*)}\alpha^{(*)}}_{ij}(\boldsymbol{\alpha},\boldsymbol{\alpha}^*)$ of the diffusion matrix. We note that constraint C3, as stated, is somewhat imprecise. Its meaning becomes clear in the implementation: each term in the evolution equation -- drift and diffusion separately -- must be built from the local Hamiltonian and its derivatives, with no dependence on $\rho$. Making this requirement fully rigorous is an open task, but the constraint suffices to determine a unique evolution equation when combined with C4--C6.

Next, we impose time-reversal invariance. We define the time-reversal operator
\begin{equation}
\mathcal{T}:(\boldsymbol{\alpha},\boldsymbol{\alpha}^*,t)\mapsto(\boldsymbol{\alpha}^*,\boldsymbol{\alpha},-t)\,,\label{time_reversal_op}
\end{equation}
corresponding to reversing momenta ($\mathbf{p}\mapsto -\mathbf{p}$, or equivalently $\boldsymbol{\alpha}\mapsto\boldsymbol{\alpha}^*$) and reversing time. We focus on Hamiltonians that are themselves invariant under momentum reversal:
\begin{equation}
H(\boldsymbol{\alpha},\boldsymbol{\alpha}^*)=H(\boldsymbol{\alpha}^*,\boldsymbol{\alpha})\,.\label{H_mom_invariant}
\end{equation}

For such Hamiltonians, the Liouville operator 
\begin{equation}
\mathcal{L}_L=i\sum_{i}
\left[
\frac{\partial}{\partial \alpha_i}
\Bigl(\frac{\partial H}{\partial \alpha_i^*}\Bigr)
- \frac{\partial}{\partial \alpha_i^*}
\Bigl(\frac{\partial H}{\partial \alpha_i}\Bigr)
\right]\label{Liouville_operator}
\end{equation}
satisfies
\begin{equation}
\mathcal{T}\mathcal{L}_L\mathcal{T}^{-1}=-\mathcal{L}_L\,,\label{Liouville_symmetry}
\end{equation}
encoding time-reversal invariance of the classical Liouville equation.

We impose time-reversal invariance on the full stochastic dynamics:
\begin{itemize}
\item \textbf{(C4) Time-reversal invariance}: For Hamiltonians satisfying Eq.\ (\ref{H_mom_invariant}), the full evolution operator $\mathcal{L} =\mathcal{L}_L+\mathcal{D}$ obeys
\begin{equation}
\mathcal{T}\mathcal{L} \mathcal{T}^{-1}=-\mathcal{L}\,.\label{full_symmetry}
\end{equation}
\end{itemize}
A comment on the motivation for C4 is in order. In classical statistical mechanics, the Boltzmann equation and other kinetic equations governing the evolution of probability distributions are time-asymmetric, even though the underlying Hamiltonian microdynamics is time-reversal invariant. One might therefore question whether time-reversal invariance should be imposed at the level of the evolution equation for $\rho$ rather than at the level of the underlying microdynamics. In principle, time-symmetric microdynamics could give rise to a time-asymmetric probability density evolution for typical boundary conditions, just as in the classical case. We take our primary motivation from the structure of the Schr\"odinger equation itself, which \emph{is} time-reversal invariant at the level of the state evolution. Imposing C4 is thus an ansatz guided by the target theory. We note, however, that relaxing C4 to require only time-reversal invariance of the microdynamics (rather than of the probability density evolution) remains an interesting direction for future investigation.

From Eqs.\ (\ref{Liouville_symmetry}) and (\ref{full_symmetry}), it follows that the diffusion operator must satisfy
\begin{equation}
\mathcal{T}\mathcal{D}\mathcal{T}^{-1}=-\mathcal{D}\,.\label{diffusion_symmetry}
\end{equation}

Applying $\mathcal{T}$ to the diffusion term in Eq.\ (\ref{diffusion_simplified}), we obtain
\begin{align}
\mathcal{T}\mathcal{D}\mathcal{T}^{-1}\rho
&=\frac{i\hbar}{2} \sum_{i,j}
\bigg\{
\frac{\partial^2}{\partial \alpha_i^* \partial \alpha_j^*}
\bigl(D_{ij}^{\alpha\alpha}(\boldsymbol{\alpha}^*,\boldsymbol{\alpha})\,\rho\bigr)\nonumber\\
&\quad+ \frac{\partial^2}{\partial \alpha_i \partial \alpha_j}
\bigl(D_{ij}^{\alpha^*\alpha^*}(\boldsymbol{\alpha}^*,\boldsymbol{\alpha})\,\rho\bigr)\nonumber\\
&\quad+ 2\frac{\partial^2}{\partial \alpha_i^* \partial \alpha_j}
\bigl(D_{ij}^{\alpha\alpha^*}(\boldsymbol{\alpha}^*,\boldsymbol{\alpha})\,\rho\bigr)
\bigg\}\!.\label{T_D_T_inv}
\end{align}

For this to equal $-\mathcal{D}\rho$, a term-by-term comparison with Eq.\ (\ref{diffusion_simplified}) yields, for any index pair $i,j$:
\begin{align}
     D_{ij}^{\alpha\alpha}(\boldsymbol{\alpha},\boldsymbol{\alpha}^*)&=
     -D_{ij}^{\alpha^{*}\alpha^{*}}(\boldsymbol{\alpha}^*,\boldsymbol{\alpha})\,,\label{TR_cond1}\\
     D_{ij}^{\alpha^*\alpha^*}(\boldsymbol{\alpha},\boldsymbol{\alpha}^*)&=
     -D_{ij}^{\alpha\alpha}(\boldsymbol{\alpha}^*,\boldsymbol{\alpha})\,,\label{TR_cond2}\\
     D_{ij}^{\alpha\alpha^*}(\boldsymbol{\alpha},\boldsymbol{\alpha}^*)&=
     -D_{ji}^{\alpha\alpha^*}(\boldsymbol{\alpha}^*,\boldsymbol{\alpha})\,.\label{TR_cond3}
\end{align}

These time-reversal conditions, combined with the reality constraints Eqs.\ (\ref{reality1})--(\ref{reality2}), determine the structure of the diffusion matrix. From Eq.\ (\ref{reality1}), we have
\begin{equation}
\left(D_{ij}^{\alpha\alpha}(\boldsymbol{\alpha},\boldsymbol{\alpha}^*)\right)^*=-D_{ij}^{\alpha^*\alpha^*}(\boldsymbol{\alpha},\boldsymbol{\alpha}^*)\,.\label{reality1_reminder}
\end{equation}
Combining this with Eq.\ (\ref{TR_cond1}) gives
\begin{equation}
\left(D_{ij}^{\alpha\alpha}(\boldsymbol{\alpha},\boldsymbol{\alpha}^*)\right)^*=D_{ij}^{\alpha\alpha}(\boldsymbol{\alpha}^*,\boldsymbol{\alpha})\,.\label{swap_is_conj}
\end{equation}
In words: for $D^{\alpha\alpha}$, swapping $\boldsymbol{\alpha} \leftrightarrow \boldsymbol{\alpha}^*$ is equivalent to complex conjugation. This is consistent with C3, since $D^{\alpha\alpha}$ is constructed from derivatives of the real-valued Hamiltonian $H(\boldsymbol{\alpha},\boldsymbol{\alpha}^*)$ that is itself invariant under $\boldsymbol{\alpha} \leftrightarrow \boldsymbol{\alpha}^*$.

For the off-diagonal block, Eq.\ (\ref{reality2}) gives
\begin{equation}
\left(D_{ij}^{\alpha\alpha^*}(\boldsymbol{\alpha},\boldsymbol{\alpha}^*)\right)^*=-D_{ji}^{\alpha^*\alpha}(\boldsymbol{\alpha},\boldsymbol{\alpha}^*)\,.
\end{equation}
Using the symmetry property $D_{ij}^{\alpha\alpha^*}=D_{ji}^{\alpha^*\alpha}$ from Eq.\ (\ref{symm3}), this becomes
\begin{equation}
\left(D_{ij}^{\alpha\alpha^*}(\boldsymbol{\alpha},\boldsymbol{\alpha}^*)\right)^*=-D_{ij}^{\alpha\alpha^*}(\boldsymbol{\alpha},\boldsymbol{\alpha}^*)\,,
\end{equation}
confirming the skew-Hermiticity established in Eq.\ (\ref{skew_hermitian}).

Specializing the time-reversal condition Eq.~(\ref{TR_cond3}) to the off-diagonal block gives
\begin{equation}
D_{ij}^{\alpha\alpha^*}(\boldsymbol{\alpha},\boldsymbol{\alpha}^*)
=
- D_{ji}^{\alpha\alpha^*}(\boldsymbol{\alpha}^*,\boldsymbol{\alpha}) \,.
\label{TR_offdiag}
\end{equation}
Combining this with the skew-Hermiticity condition Eq.~(\ref{skew_hermitian}), we obtain
\begin{equation}
D_{ij}^{\alpha\alpha^*}(\boldsymbol{\alpha},\boldsymbol{\alpha}^*)
=
\left(D_{ij}^{\alpha\alpha^*}(\boldsymbol{\alpha}^*,\boldsymbol{\alpha})\right)^* .
\label{TR_conj_offdiag}
\end{equation}
Thus, in the off-diagonal block, swapping $\alpha \leftrightarrow \alpha^*$ is equivalent both to transposing the mode indices and to complex conjugation. We now show that constraint C3 then forces $D_{ij}^{\alpha\alpha^*}=0$. Any expression built from derivatives of the real, swap-invariant Hamiltonian $H(\alpha,\alpha^*)$ without explicit factors of $i$ obeys the property that complex conjugation is implemented by the swap $\alpha \leftrightarrow \alpha^*$. For mixed derivatives, this swap sends $\partial^2 H/\partial\alpha_i \partial\alpha_j^*$ to $\partial^2 H/\partial\alpha_i^* \partial\alpha_j$, which by equality of mixed partial derivatives equals $\partial^2 H/\partial\alpha_j \partial\alpha_i^*$. Hence, in the mixed sector, conjugation acts as transposition, so constructions from derivatives of $H$ without explicit factors of $i$ are necessarily symmetric in $i,j$. This is incompatible with the skew-Hermitian character of $D_{ij}^{\alpha\alpha^*}$, and therefore
\begin{equation}
D_{ij}^{\alpha\alpha^*}(\boldsymbol{\alpha},\boldsymbol{\alpha}^*) = 0 \,.
\label{offdiag_vanishes}
\end{equation}
An important consequence of these time-reversal conditions, as we shall see in Section~\ref{subsec:traceless}, is that the diffusion matrix, when expressed in real phase-space coordinates $(q_i, p_i)$, is traceless. This tracelessness -- encoding equal numbers of positive and negative eigenvalues -- is the key feature distinguishing time-reversal-invariant stochastic dynamics from standard diffusion processes. We return to this point in Section~\ref{subsec:traceless} after deriving the explicit form of the diffusion matrix.

\subsection{Energy conservation and minimality}

Having established that time-reversal invariance requires a diffusion matrix with vanishing off-diagonal blocks $D_{ij}^{\alpha\alpha^*}$, we now impose energy conservation to further constrain its form.

\begin{itemize}
\item \textbf{(C5) Energy conservation on expectation}: The expectation value of the Hamiltonian is conserved under time evolution:
\begin{equation}
\frac{d}{dt}\langle H\rangle = 0\,.\label{energy_conservation}
\end{equation}
\end{itemize}

Hamilton's equations conserve energy, so the Liouville term in Eq.\ (\ref{general_Fokker_Planck}) already satisfies this constraint. We therefore need only ensure that the diffusion term does not violate energy conservation. The time derivative of the energy expectation is
\begin{align}
    \frac{d}{dt}\langle H\rangle&=\int H(\mathbf{z})\,\frac{\partial\rho(\mathbf{z},t)}{\partial t}\, d\mathbf{z}\nonumber\\
    &=\int H(\mathbf{z})\,(\mathcal{L}_L+\mathcal{D})\rho(\mathbf{z},t)\, d\mathbf{z}\nonumber\\
    &=\int H(\mathbf{z})\,\mathcal{D}\rho(\mathbf{z},t)\, d\mathbf{z}\,,\label{energy_deriv}
\end{align}
where the last equality uses energy conservation under classical Hamiltonian evolution, i.e., $\int H(\mathbf{z})\,\mathcal{L}_L\rho\, d\mathbf{z} = 0$.

Substituting the diffusion operator from Eq.\ (\ref{diffusion_simplified}) with $D_{ij}^{\alpha\alpha^*} = 0$:
\begin{align}\label{after_integration_by_parts}
    &\frac{d}{dt}\langle H\rangle
    =\int H(\mathbf{z})\frac{i\hbar}{2} \\&\hspace{0.5cm}\sum_{i,j}
\left\{
\frac{\partial^2}{\partial \alpha_i \partial \alpha_j}
\bigl(D_{ij}^{\alpha\alpha}\,\rho\bigr)
- \frac{\partial^2}{\partial \alpha_i^* \partial \alpha_j^*}
\bigl((D_{ij}^{\alpha\alpha})^*\,\rho\bigr)
\right\} d\mathbf{z}\nonumber\\
    &=\frac{i\hbar}{2} \sum_{i,j}\int
\left\{
\frac{\partial^2H(\mathbf{z})}{\partial \alpha_i \partial \alpha_j}
D_{ij}^{\alpha\alpha}
- \frac{\partial^2H(\mathbf{z})}{\partial \alpha_i^* \partial \alpha_j^*}
(D_{ij}^{\alpha\alpha})^*
\right\}\,\rho\, d\mathbf{z}\,.\nonumber
\end{align}
In the last step, we integrated by parts twice with respect to each pair of variables $(\alpha_i, \alpha_j)$ or $(\alpha_i^*, \alpha_j^*)$, assuming that $\rho$ and its derivatives vanish sufficiently rapidly at infinity for boundary terms to vanish.

For energy to be conserved for arbitrary probability densities $\rho(\mathbf{z},t)$, the integrand must vanish identically:
\begin{equation}
\frac{\partial^2H(\mathbf{z})}{\partial \alpha_i \partial \alpha_j}D_{ij}^{\alpha\alpha}
- \frac{\partial^2H(\mathbf{z})}{\partial \alpha_i^* \partial \alpha_j^*}
(D_{ij}^{\alpha\alpha})^*=0\,.\label{energy_cons_condition}
\end{equation}

This condition is satisfied if
\begin{equation}
D_{ij}^{\alpha\alpha}
= f_{ij}(\boldsymbol{\alpha},\boldsymbol{\alpha}^*)\,\frac{\partial^2H}{\partial \alpha_i^* \partial \alpha_j^*}\,,\label{D_proportional_to_Hessian}
\end{equation}
for some function $f_{ij}(\boldsymbol{\alpha},\boldsymbol{\alpha}^*)$ that must be real-valued (since both $D_{ij}^{\alpha\alpha}$ and $\frac{\partial^2H}{\partial \alpha_i^* \partial \alpha_j^*}$ transform identically under complex conjugation).

Constraint C3 (local Hamiltonian dependence) allows $f_{ij}$ to depend on $(\boldsymbol{\alpha},\boldsymbol{\alpha}^*)$ only through the Hamiltonian and its derivatives. The most general form consistent with C3 would be
\begin{equation}
f_{ij}(\boldsymbol{\alpha},\boldsymbol{\alpha}^*) = \sum_{k,\ell,m,n,\ldots} c_{ij}^{k\ell mn\ldots} \frac{\partial^k H}{\partial \alpha_\ell}\frac{\partial^m H}{\partial \alpha_n^*}\cdots\,,
\end{equation}
involving products of various derivatives of $H$ with coefficient tensors $c_{ij}^{k\ell mn\ldots}$.

To uniquely determine the theory, we invoke a final principle:
\begin{itemize}
\item \textbf{(C6) Minimality}: The diffusion matrix contains no terms or dependencies beyond those strictly required by constraints C1--C5.
\end{itemize}

This principle of parsimony requires that $f_{ij}$ be independent of $(\boldsymbol{\alpha},\boldsymbol{\alpha}^*)$ -- i.e., that it be a constant:
\begin{equation}
f_{ij}(\boldsymbol{\alpha},\boldsymbol{\alpha}^*) = C\,\delta_{ij}\,,\label{f_is_constant}
\end{equation}
where $C$ is a global constant and we have also used the symmetry $f_{ij} = f_{ji}$ (since $D^{\alpha\alpha}$ is symmetric) to write $f_{ij} = C\,\delta_{ij}$ for some scalar $C$.

\subsection{The final equation in complex phase space variables}

We can now state the main result of this section:

\noindent\textbf{Main Result.}
Let the time evolution of a probability density $\rho(\mathbf{z},t)$ on phase space satisfy:
\begin{enumerate}[label=\textup{(C\arabic*)},leftmargin=2.5em]
    \item \textbf{Classical limit}: The evolution consists of a Liouville term plus a stochasticity term that vanishes as $\hbar \to 0$.
    \item \textbf{Fokker-Planck form}: The evolution takes Fokker-Planck form with a symmetric diffusion matrix $D_{ij}$, with no constraint on the signs of its eigenvalues.
    \item \textbf{Local Hamiltonian dependence}: All parameters in the evolution equation are constructed from the local Hamiltonian $H(\boldsymbol{\alpha},\boldsymbol{\alpha}^*)$ and its derivatives.
    \item \textbf{Time-reversal invariance}: For Hamiltonians satisfying $H(\boldsymbol{\alpha},\boldsymbol{\alpha}^*)=H(\boldsymbol{\alpha}^*,\boldsymbol{\alpha})$, the evolution operator $\mathcal{L}$ obeys $\mathcal{T}\mathcal{L}\mathcal{T}^{-1}=-\mathcal{L}$.
    \item \textbf{Energy conservation}: $\frac{d}{dt}\langle H\rangle = 0$.
    \item \textbf{Minimality}: The diffusion matrix contains no terms beyond those required by C1--C5.
\end{enumerate}
Then the evolution equation must take the form
\begin{widetext}
\begin{align}
\frac{\partial \rho}{\partial t}
&= i\sum_{i}
\left[
\frac{\partial}{\partial \alpha_i}
\Bigl(\frac{\partial H}{\partial \alpha_i^*}\,\rho\Bigr)
- \frac{\partial}{\partial \alpha_i^*}
\Bigl(\frac{\partial H}{\partial \alpha_i}\,\rho\Bigr)
\right]
\label{final_FP}\\[4pt]
&\quad
+ \frac{iC\hbar}{2} \sum_{i,j}
\left\{
\frac{\partial^2}{\partial \alpha_i \partial \alpha_j}
\Bigl(\frac{\partial^2 H}{\partial \alpha_i^*\partial \alpha_j^*}\,\rho\Bigr)
- \frac{\partial^2}{\partial \alpha_i^* \partial \alpha_j^*}
\Bigl(\frac{\partial^2 H}{\partial \alpha_i\partial \alpha_j}\,\rho\Bigr)
\right\}\!.\nonumber
\end{align}
\end{widetext}
for some constant $C$.

The derivation occupies the preceding subsections. Eq.~(\ref{final_FP}) is the unique time-reversal-invariant stochastic generalization of the Liouville equation satisfying our constraints. As we demonstrate in Section~\ref{sec:Drummond}, it coincides exactly with the evolution of the Husimi function in bosonic quantum field theory, provided that $C$ and the scales $\ell_i$ are appropriately chosen.

\subsection{Real phase-space formulation and tracelessness}\label{subsec:traceless}

Having derived the evolution equation in complex phase-space coordinates $(\boldsymbol{\alpha}, \boldsymbol{\alpha}^*)$, we now transform to real phase-space variables $(q_j, p_j)$ to make explicit two key features: the tracelessness of the diffusion matrix and its connection to the Hessian of the Hamiltonian in real variables.

\subsubsection{Transformation to real variables}

We work with the real phase-space coordinate vector
\begin{equation}
\xi = \begin{pmatrix}
q_1\\ \vdots\\ q_N\\ p_1\\ \vdots\\ p_N
\end{pmatrix}\!,\label{real_phase_space}
\end{equation}
where $q_j$ and $p_j$ are canonical position and momentum coordinates. In real variables, the classical Liouville equation takes the standard Hamiltonian form
\begin{equation}
\frac{\partial \rho(\xi,t)}{\partial t}
= \{H,\rho\}_{\mathrm{PB}} = \sum_{j=1}^N\left(\frac{\partial H}{\partial q_j}\,\frac{\partial \rho}{\partial p_j} - \frac{\partial H}{\partial p_j}\,\frac{\partial \rho}{\partial q_j}\right)\!,\label{Liouville_real}
\end{equation}
where $\{\cdot,\cdot\}_{\mathrm{PB}}$ denotes the Poisson bracket.

To connect with our complex-variable derivation, we introduce dimensionless scaled coordinates. Define
\begin{equation}
Q_j := \frac{q_j}{\ell_j}, \qquad P_j := \frac{p_j\,\ell_j}{\kappa}, \qquad z = \begin{pmatrix}
Q_1\\ \vdots\\ Q_N\\ P_1\\ \vdots\\ P_N
\end{pmatrix}\!,\label{dimensionless_coords}
\end{equation}
with $\ell_j$ and $\kappa$ as introduced before. The transformation between dimensionful and dimensionless variables is given by the diagonal scaling matrix
\begin{equation}
L = \mathrm{diag}\!\left(\ell_1,\dots,\ell_N,\,\frac{\kappa}{\ell_1},\dots,\frac{\kappa}{\ell_N}\right)\!,
\quad \text{so that} \quad \xi = Lz.\label{scaling_matrix}
\end{equation}

The complex variables $\alpha_j$ defined in Eq.\ (\ref{complex_variable}) relate to these dimensionless real coordinates via
\begin{equation}
\alpha_j = \sqrt{\frac{\kappa}{2}}(Q_j + iP_j), \qquad \alpha_j^* = \sqrt\frac{\kappa}{2}(Q_j - iP_j)\,.\label{complex_to_real}
\end{equation}

\subsubsection{The diffusion matrix in real variables}

The Hessian of the Hamiltonian in physical variables is the $2N \times 2N$ matrix
\begin{equation}
H''(\xi) = \frac{\partial^2 H(\xi)}{\partial \xi\,\partial \xi^{\mathsf{T}}}\!.\label{physical_Hessian}
\end{equation}
Transforming to dimensionless variables yields the scaled Hessian
\begin{equation}
\tilde{H}''(z) := \frac{\partial^2 H(Lz)}{\partial z\,\partial z^{\mathsf{T}}} = L^{\mathsf{T}} H''(\xi)\,L\,.\label{scaled_Hessian}
\end{equation}

To express the diffusion matrix in real variables, we also introduce the standard symplectic matrix
\begin{equation}
J = \begin{pmatrix}
0 & I_N\\ -I_N & 0
\end{pmatrix}\!,\label{symplectic_matrix}
\end{equation}
where $I_N$ is the $N \times N$ identity matrix.

Using this matrix, the diffusion matrix derived in complex variables can be elegantly written in real coordinates. As shown in Appendix~\ref{app:complex_to_real}, the result takes the commutator form
\begin{equation}
D(z) = \frac{C}{2\kappa^2}\bigl[\tilde{H}''(z),\ J\bigr]\,,
\label{eq:diffusion_commutator}
\end{equation}
where $J$ is the symplectic matrix Eq.\ (\ref{symplectic_matrix}) and $\tilde{H}''(z)$ is the scaled Hessian Eq.\ (\ref{scaled_Hessian}).

\subsubsection{Tracelessness}

This commutator form immediately reveals that the diffusion matrix is \emph{traceless}:
\begin{align}
\mathrm{Tr}\bigl(D(z)\bigr) &= \frac{C}{2}\,\mathrm{Tr}\bigl(\tilde{H}''(z)\,J - J\,\tilde{H}''(z)\bigr)\nonumber\\
&= \frac{C}{2}\,\bigl(\mathrm{Tr}(\tilde{H}''(z)\,J)-\mathrm{Tr}(J\,\tilde{H}''(z)) \bigr)\nonumber\\
&= 0\,,\label{traceless_proof}
\end{align}
where the last equality follows from the cyclicity of the trace: $\mathrm{Tr}(AB) = \mathrm{Tr}(BA)$.

This tracelessness is the defining feature of time-reversal-invariant stochastic dynamics. Unlike standard diffusion processes where all eigenvalues of the diffusion matrix are positive (corresponding to spreading in all directions), a traceless diffusion matrix has equal numbers of positive and negative eigenvalues. 
This can be seen directly from the commutator structure of the diffusion matrix in real coordinates: it anticommutes with the symplectic matrix $J$, so whenever $v$ is an eigenvector with eigenvalue $\lambda$, $Jv$ is an eigenvector with eigenvalue $-\lambda$. Hence the nonzero eigenvalues come in positive-negative pairs of equal multiplicity, with any remainder consisting only of zero modes. The positive eigenvalues correspond to forward-time diffusion along certain directions in phase space, while the negative eigenvalues correspond to backward-time diffusion along complementary directions. This balanced structure -- enforced by constraint C4 -- is what allows the dynamics to treat both time directions symmetrically.

\subsubsection{Complete evolution equation in real variables}

To write the full Fokker-Planck equation in real variables, we define the dimensionless gradient operator
\begin{equation}
\tilde{\nabla} := \begin{pmatrix}
\ell_1\,\dfrac{\partial}{\partial q_1}\\ \vdots\\ \ell_N\,\dfrac{\partial}{\partial q_N}\\ \dfrac{\kappa}{\ell_1}\,\dfrac{\partial}{\partial p_1}\\ \vdots\\ \dfrac{\kappa}{\ell_N}\,\dfrac{\partial}{\partial p_N}
\end{pmatrix} = \begin{pmatrix}
\tilde{\partial}_1\\ \vdots\\ \tilde{\partial}_{2N}
\end{pmatrix}\!.\label{scaled_gradient}
\end{equation}

The diffusion contribution to the time evolution can then be written compactly as
\begin{equation}
\bigl(\partial_t \rho\bigr)_{\mathrm{diff}} = \frac{\hbar C}{4\kappa^2}\sum_{a,b=1}^{2N}\tilde{\partial}_a \tilde{\partial}_b\Bigl(\bigl[\tilde{H}'',\,J\bigr]_{ab}\,\rho\Bigr)\,.\label{diffusion_components}
\end{equation}

Combining drift and diffusion, the complete time evolution equation is:
\begin{equation}
\boxed{
\begin{aligned}
\frac{\partial \rho}{\partial t} 
&= \sum_{j=1}^N\left(\frac{\partial H}{\partial q_j}\,\frac{\partial \rho}{\partial p_j} - \frac{\partial H}{\partial p_j}\,\frac{\partial \rho}{\partial q_j}\right) \\
&+\quad \frac{\hbar C}{4\kappa^2}\sum_{a,b=1}^{2N}\tilde{\partial}_a \tilde{\partial}_b\Bigl(\bigl[\tilde{H}'',\,J\bigr]_{ab}\,\rho\Bigr)\,.
\end{aligned}
}
\label{eq:full_FPE_real}
\end{equation}
This equation reveals the structure of our time-symmetric stochastic generalization: classical Hamiltonian flow (first line) plus traceless diffusion determined by the commutator of the symplectic matrix with the Hessian (second line). The constant $C$ remains to be fixed by comparison with quantum theory, which we undertake in Section~\ref{sec:Drummond}.

\subsection{Fixing the length scales}\label{subsec:scale_fixing}

The length scales $\ell_i$ appearing in the definition of the complex phase-space variables, Eq.~(\ref{complex_variable}), remain undetermined by constraints C1--C6. Different choices lead to different diffusion terms for the same Hamiltonian. A natural choice that uniquely fixes this remaining freedom is:
\begin{itemize}
\item \textbf{(C7) Classical free dynamics}: For non-interacting (quadratic) Hamiltonians, the diffusion term vanishes, and the evolution reduces to the classical Liouville evolution.
\end{itemize}

As we will see below, this requirement yields the complex phase space variables that, after quantization, correspond to the standard creation and annihilation operators for the harmonic oscillator. Moreover, it aligns two distinctions: deterministic versus stochastic dynamics, and free versus interacting theories. When C7 is implemented, free field theories -- described by quadratic Hamiltonians -- evolve in accordance with the Liouville equation. This suggests a perspective according to which stochastic behaviour arises only from interactions. One may see this as physically natural: in the absence of interactions, there is nothing to generate the fluctuations that distinguish quantum from classical behaviour.

One might object here, with an eye on the envisaged recovery of quantum field theories from this approach, that free quantum systems such as the harmonic oscillator already exhibit non-classical features, most notably discrete energy spectra. Indeed, if one implements C7, the discreteness of energy spectra must be viewed not as a property of the free field evolution itself but arises from the interaction involved in probing the system -- for instance, the coupling between photon and electron modes in spectroscopic measurements. For a satisfying account of how measurement outcomes are generated, a detailed investigation of these interactions in terms of whatever microdynamics may (hopefully!) be found to underlie Eq.~(\ref{eq:full_FPE_real}) would certainly be needed. Further below we offer some tentative consideration of how discrete measurement outcomes might be generated when discussing empirical adequacy. In any case, as we will see now, the alignment of the free/interacting and deterministic/stochastic distinctions is naturally combined with the previous assumptions when one uses the standard harmonic oscillator length in the free case and keeps it in the presence of interactions.

Constraint C7 requires that $[\tilde{H}''_0,\,J] = 0$ for the free Hamiltonian $H_0$. Since the diffusion matrix is the commutator of the symplectic matrix with the scaled Hessian, Eq.~(\ref{eq:diffusion_commutator}), this condition is satisfied if and only if $\tilde{H}''_0$ is proportional to the identity matrix.

For a free scalar field, the Hamiltonian in momentum modes takes the form
\begin{equation}
H_0 = \sum_{\mathbf{k}} \omega_{\mathbf{k}} \left( \frac{p_{\mathbf{k}}^2}{2\omega_{\mathbf{k}}} + \frac{\omega_{\mathbf{k}} q_{\mathbf{k}}^2}{2} \right) = \sum_{\mathbf{k}} \omega_{\mathbf{k}} |\alpha_{\mathbf{k}}|^2\,,
\label{eq:free_hamiltonian_modes}
\end{equation}
where $\mathbf{k}$ labels momentum modes, $\omega_{\mathbf{k}} = \sqrt{\mathbf{k}^2 + m^2}$ is the mode frequency, where $m$ is the field mass, and the second equality uses the complex variables defined in Eq.~(\ref{complex_variable}). For a complex field, there is an analogous term for the antiparticle modes.

In momentum modes, the Hessian of $H_0$ is diagonal. The physical Hessian has entries
\begin{equation}
\frac{\partial^2 H_0}{\partial q_{\mathbf{k}}^2} = \omega_{\mathbf{k}}^2\,, \qquad \frac{\partial^2 H_0}{\partial p_{\mathbf{k}}^2} = 1\,,
\label{eq:free_hessian_entries}
\end{equation}
with vanishing off-diagonal and mixed $q$--$p$ components. The scaled Hessian, Eq.~(\ref{scaled_Hessian}), then has diagonal entries $\ell_{\mathbf{k}}^2 \omega_{\mathbf{k}}^2$ in the field sector and $\ell_{\mathbf{k}}^{-2}\kappa^2$ in the momentum sector. For these to be equal -- and thus for $\tilde{H}''_0$ to be proportional to the identity -- we require
\begin{equation}
\ell_{\mathbf{k}}^2 \omega_{\mathbf{k}}^2 = \frac{\kappa^2}{\ell_{\mathbf{k}}^2}\,, \qquad \text{i.e.,} \qquad \ell_{\mathbf{k}} = \sqrt{\frac{\kappa}{\omega_{\mathbf{k}}}}\,,
\label{eq:length_scale_fixing}
\end{equation}
which, as we show in Section~\ref{sec:Drummond}, corresponds to the characteristic mode scales in quantum field theory when $\kappa$ is identified with $\hbar$.

This construction requires working in momentum modes -- the eigenmodes of the free Hamiltonian. In position space, the free Hamiltonian includes the spatial gradient term, which couples neighbouring lattice sites. The resulting Hessian is not diagonal: it contains off-diagonal entries from the discrete Laplacian. No choice of site-dependent length scales $\ell_j$ can eliminate these off-diagonal couplings to make $\tilde{H}''_0$ proportional to the identity. For the real scalar field the condition $[J, \tilde{H}''_0] = 0$ thus uniquely selects both the mode basis (momentum modes) and the length scales ($\ell_{\mathbf{k}} \propto \omega_{\mathbf{k}}^{-1/2}$).

When interactions are present, we retain the length scales $\ell_{\mathbf{k}}$ at their free-theory values, Eq.~(\ref{eq:length_scale_fixing}). This ensures that the free part of the Hamiltonian continues to contribute only to the drift term, not to the diffusion. The interaction Hamiltonian $H_{\mathrm{int}}$ -- which has a field-dependent Hessian that is not proportional to the identity -- then generates the diffusion term through $[ \tilde{H}''_{\mathrm{int}},\,J]$.

This prescription has the following physical rationale. Consider a scattering problem, with individual mode excitations propagating freely towards and away from a localized interaction region. It seems plausible to require that in the asymptotic regions -- far from the interaction, in the distant past and distant future -- the canonical field momenta $\pi_{\mathbf{k}}$ should correspond directly to the kinetic momenta, i.e., to the actual velocities $\dot{\phi}_{\mathbf{k}}^*$ of the field modes.

If the length scales $\ell_{\mathbf{k}}$ differed from their free-theory values, the free Hamiltonian would contribute to the diffusion term even in these asymptotic regions. Canonical momenta would then no longer track velocities, and the clean separation between free propagation and interaction -- which underlies the standard definition of scattering states -- would be lost.

Thus, fixing $\ell_{\mathbf{k}} = \sqrt{\kappa/\omega_{\mathbf{k}}}$ ensures that stochastic behaviour arises solely from interactions, while asymptotic states retain their classical character. It should be noted, however, that the alignment between the distinctions deterministic/stochastic and free/interacting theories that is implemented via C7 is not entailed by C1--C6 and could in principle be given up without abandoning these.

Finally, we note that the length scales $\ell_{\mathbf{k}} = \sqrt{\kappa/\omega_{\mathbf{k}}}$ admit a natural Lorentz-covariant generalization as the operator $\hat{L} = \kappa^{1/2}(-\Box + m^2)^{-1/4}$, where $\Box = \partial_\mu \partial^\mu$ is the d'Alembertian. This suggests that the present framework may extend to relativistic quantum field theories while preserving constraint C7. We leave the investigation of Lorentz covariance to future work.

We have now derived the unique form of the time-symmetric stochastic evolution equation satisfying constraints C1--C7. One free parameter remains: the global constant $C$. In the following section, we determine both by requiring agreement with quantum field theory, which will fix the value of $C$.

\section{Connection to quantum field theory}
\label{sec:Drummond}

We now demonstrate that the evolution equation derived in Section~\ref{sec:derivation} coincides exactly with the time evolution of the Husimi function in bosonic quantum field theory. This unexpected agreement will be used to suggest an interpretation of quantum field theory as the statistical mechanics of a time-symmetric stochastic theory.

\subsection{The Husimi function}

In quantum theory, the Husimi Q-function \citep{husimi1940} is a phase-space representation of the density matrix $\hat{\rho}$ defined by
\begin{equation}
    Q(\boldsymbol{\alpha}, \boldsymbol{\alpha}^*,t)=\frac{1}{\pi^N}\bra{\boldsymbol{\alpha}}\hat{\rho}(\hat{a}, \hat{a}^\dagger,t)\ket{\boldsymbol{\alpha}}\,,\label{husimi}
\end{equation}
where $\ket{\boldsymbol{\alpha}}$ are coherent states labeled by complex phase-space coordinates $\boldsymbol{\alpha}$. Alongside the Wigner function and the Glauber Sudarshan P-function, the Husimi Q-function is among the most frequently used ``quasiprobability distributions'' that express the quantum state in phase space. Each of these functions can be used to calculate quantum expectation values using phase space integrals. For the Husimi function, this can be done with the formula
\begin{equation}
    \mathrm{Tr}(\hat{A}\hat{\rho})=\int A(\boldsymbol{\alpha})\,Q(\boldsymbol{\alpha})\,d\boldsymbol{\alpha}\,,\label{expectation}
\end{equation}
provided $A(\boldsymbol{\alpha})$ is the phase space function mapped to $\hat A$ by Anti-Wick quantization. (If the Wigner function is used instead of the Q-function, one must use the phase space function mapped to $\hat A$ by Weyl quantization. If the Glauber-Sudarshan function is used, one must use the phase space function mapped to $\hat A$ by Wick quantization. See \citep{lee,folland1989} for details.)

Unlike the Wigner and Glauber-Sudarshan functions, the Husimi function has the formal features of a probability density, namely, it is everywhere non-negative and normalized: $\int Q(\boldsymbol{\alpha})\,d\boldsymbol{\alpha} = 1$. Interestingly, it equals the squared modulus of the wave function on phase space (rather than position or momentum space) \citep{torresvegafrederick}.

\subsection{Anti-Wick quantization}

As noted, the relationship Eq.\ (\ref{expectation}) holds when the operator $\hat{A}$ and the phase-space function $A(\boldsymbol{\alpha})$ are connected by \emph{Anti-Wick quantization} (also called coherent-state or Berezin-Toeplitz quantization):
\begin{equation}
    A\mapsto\hat{A}=\frac{1}{\pi^N}\int A(\boldsymbol{\alpha})\ket{\boldsymbol{\alpha}}\bra{\boldsymbol{\alpha}}d\boldsymbol{\alpha}\,.\label{Anti-Wick}
\end{equation}

This quantization scheme differs from the more commonly used Weyl quantization. (See \citep[Ch.\ 13]{hall2013} for a useful comparison.) For the Hamiltonian operator $\hat{H}$, we denote its Anti-Wick symbol -- the phase-space function satisfying Eq.\ (\ref{Anti-Wick}) -- by $H_{\mathrm{aW}}(\boldsymbol{\alpha}, \boldsymbol{\alpha}^*)$. Incidentally, this is the Hamiltonian symbol that appears in the phase space path integral formulation of quantum mechanics \citep{daubechiesklauder1985}. For field theories on continuous space, Anti-Wick quantization is generally non-unique: it requires a choice of one-particle structure (equivalently, a choice of vacuum state), and different choices give rise to different Husimi functions and potentially different stochastic evolutions. This non-uniqueness is well understood in the algebraic approach to quantum field theory \citep{browning2020}, where it is related to the non-uniqueness of the Segal-Bargmann (complex wave) representation \citep[Ch.~1.11]{baezsegalzhou}. Since our analysis is carried out on a lattice with finitely many degrees of freedom, this issue does not arise here. However, it represents an important challenge for any future rigorous extension to the continuum limit.

\subsection{Time evolution of the Husimi function}

The time evolution of $Q$ is obtained as the Schr\"odinger (or von Neumann) equation in the coherent-state representation. The fact that its leading terms correspond to a Fokker-Planck equation with a traceless diffusion matrix was pointed out by Drummond \citep{drummond2021}. However, in Drummond's calculation the parameters in the Fokker-Planck equation are expressed in terms of the quantum Hamiltonian operator, not in terms of any phase space Hamiltonian, so that equation cannot be directly compared with the equation derived in Section~\ref{sec:derivation}. In \citep{MIZRAHI1984241,MIZRAHI1986237,Appleby_2000} the evolution equation for the Husimi function was derived in terms of the \emph{Wick-} (\emph{not} Anti-Wick-) Hamiltonian symbol
\[
\bra{\boldsymbol{\alpha}}\hat{H}(\hat{a}, \hat{a}^\dagger)\ket{\boldsymbol{\alpha}}\,,
\]
i.e. the phase space function from which $\hat{H}$ is obtained by Wick quantization. However, as noted in connection with Eq.\ (\ref{expectation}), the phase space function that reproduces the expectation value of $\hat H$ when performing a phase space integral weighted with the Husimi function is actually the \emph{Anti-Wick}, not the Wick, symbol of the Hamiltonian $H_{\mathrm{aW}}$.

In terms of $H_{\mathrm{aW}}$ the general evolution equation (see \citep{tyagifriederich} for the derivation and for the definition of the multi-indices $\mathbf m$) is:
\begin{equation}
    \partial_t Q 
= \frac{i}{\hbar}\sum_{|\mathbf{m}|\ge1}\frac{\hbar^{|\mathbf{m}|}}{\mathbf{m}!}
\Big[
\partial^{\mathbf{m}}\!\bigl(\bar{\partial}^{\mathbf{m}} H_{\mathrm{aW}}\cdot Q\bigr)
-
\bar{\partial}^{\mathbf{m}}\!\bigl(\partial^{\mathbf{m}} H_{\mathrm{aW}}\cdot Q\bigr)
\Big]\,.\label{full_Q_evolution}
\end{equation}
The key observation is that this infinite series \emph{truncates} when the Hamiltonian has limited polynomial degree. Specifically, if $H_{\mathrm{aW}}$ is a polynomial with maximum degree $d$ in each complex variable $\alpha_i$ (and its conjugate), then only terms with $|\mathbf{m}| \le d$ contribute. This follows because $\partial^{\mathbf{m}} H_{\mathrm{aW}} = 0$ whenever any component $m_i > d$.

For Hamiltonians that are polynomials which satisfy, for all $i$, $j$, $k$,
\begin{equation}
\frac{\partial^3 H_{\mathrm{aW}}}{\partial \alpha_i\partial \alpha_j\partial \alpha_k} =\frac{\partial^3 H_{\mathrm{aW}}}{\partial (\alpha_i^{*})\partial (\alpha_j^{*})\partial (\alpha_k^{*})} = 0\,,\label{condition}
\end{equation}
the series truncates at second order, yielding:
\begin{widetext}
\begin{align}
\frac{\partial Q}{\partial t}
&= i\sum_{i}
\left[
\frac{\partial}{\partial \alpha_i}
\Bigl(\frac{\partial H_{\mathrm{aW}}}{\partial \alpha_i^*}\,Q\Bigr)
- \frac{\partial}{\partial \alpha_i^*}
\Bigl(\frac{\partial H_{\mathrm{aW}}}{\partial \alpha_i}\,Q\Bigr)
\right]
\label{Q_evolution_quadratic}\\[4pt]
&\quad
+ \frac{i\hbar}{2} \sum_{i,j}
\left\{
\frac{\partial^2}{\partial \alpha_i \partial \alpha_j}
\Bigl(\frac{\partial^2 H_{\mathrm{aW}}}{\partial \alpha_i^*\partial \alpha_j^*}\,Q\Bigr)
- \frac{\partial^2}{\partial \alpha_i^* \partial \alpha_j^*}
\Bigl(\frac{\partial^2 H_{\mathrm{aW}}}{\partial \alpha_i\partial \alpha_j}\,Q\Bigr)
\right\}\!.\nonumber
\end{align}
\end{widetext}
As said, this is simply the Schr\"odinger equation, expressed in phase-space coordinates, via the coherent state representation, for this class of Hamiltonians.

\subsection{The remarkable coincidence}

Comparing Eq.\ (\ref{Q_evolution_quadratic}) with our derived equation (Eq.\ \ref{final_FP}), we observe that they have exactly the same structure! The equations are identical if we identify:
\begin{equation}
\boxed{H = H_{\mathrm{aW}}\,, \quad C=1\,.}
\label{parameter_identifications}
\end{equation}

This is a striking result: The simplest stochastic generalization of classical mechanics satisfying constraints C1--C7 coincides precisely with the Husimi evolution in bosonic quantum field theories with Hamiltonians that fulfil Eq.~(\ref{condition}). This correspondence not only fixes the free parameter ($C = 1$) but also vindicates the minimality assumption (C6) retrospectively: had we chosen a more complicated form for the diffusion matrix, the match with quantum theory would have been lost.

\subsection{The probability interpretation}

This exact correspondence suggests a natural interpretation: In those quantum field theories where the Husimi evolution truncates at Fokker-Planck form, the Husimi function can be understood as a genuine probability density $\rho$ on phase space, rather than merely a quasiprobability distribution.

Under this interpretation:
\begin{itemize}
\item At each instant, quantum systems have definite locations in phase space.
\item All dynamical variables -- the phase-space functions linked to quantum operators via Anti-Wick quantization -- have sharp values at each time.
\item The Husimi function, analogous to probability densities in classical statistical mechanics, reflects incomplete information about the system's instantaneous phase-space location (or, alternatively, describes an ensemble of systems rather than an individual system).
\item Quantum statistics arise from this epistemic uncertainty, rather than from fundamental indefiniteness of observables.
\end{itemize}

The idea of interpreting the Husimi $Q$-function as a proper probability density was proposed by Bopp \citep{bopp1956} and has been recently revived from physical \citep{drummondreid2020} and philosophical \citep{friederich2024} perspectives. It underlies the phase-space formulation of Bohmian mechanics \citep{depolavieja1996}, which treats position and momentum democratically. The Husimi function provides correct measurement statistics in heterodyne detection, which simultaneously probes conjugate quadratures \citep{PhysRevLett.117.070801,Wiseman_Milburn_2009}, and Appleby showed that it yields the distribution of results in retrodictively optimal phase-space measurements \citep{appleby1998optimaljointmeasurementsposition}.

The present work provides new support for interpreting the Husimi function as a proper probability density by showing that the Husimi function evolution follows naturally from time-reversal invariance and energy conservation -- fundamental physical principles that make no reference to quantum theory. However, establishing that this probability density can be understood as arising from an ensemble of underlying stochastic trajectories requires additional analysis of the micro-dynamics, which we undertake in a companion paper. We now turn to discussing the limitations of the present framework by discussing to which quantum field theories it applies.

\section{Applicability}
\label{sec:standard_model}

The evolution equation derived in Section~\ref{sec:derivation} reproduces the Husimi function dynamics in quantum field theories whose Anti-Wick Hamiltonian symbol satisfies the truncation condition Eq.~(\ref{condition}). This class includes all free bosonic field theories. It also includes interacting theories with quartic \emph{density-density} couplings in the complex phase space variables, where the interaction at each mode or lattice site depends only on the local density $|\alpha_i|^2$ at that site. However, generic interaction terms in the Standard Model of particle physics lead to Hamiltonians that violate the truncation condition, taking the Husimi evolution beyond Fokker-Planck form.

In this section, we first identify an important interacting model that is covered by our framework (Section~\ref{subsec:bose_hubbard}), then examine Standard Model terms that are not (Sections~\ref{subsec:higgs} and \ref{subsec:gauge}), and finally discuss possible reactions to these limitations (Section~\ref{subsec:reactions}).

\subsection{The Bose-Hubbard model}\label{subsec:bose_hubbard}

The Bose-Hubbard model describes ultracold bosonic atoms in optical lattices and exhibits a quantum phase transition between superfluid and Mott insulator phases \citep{fisher1989,jaksch1998,greiner2002,bloch2008}. We use this model to illustrate how one obtains the phase space Hamiltonian (in the form of an Anti-Wick symbol) from the Hamiltonian operator, which is
\begin{equation}
\hat{H}_{\mathrm{BH}} = -J\sum_{\langle i,j\rangle}\hat{a}_i^\dagger\hat{a}_j + \frac{U}{2}\sum_i\hat{n}_i(\hat{n}_i-1)\,,
\end{equation}
where $\hat{a}_i$ and $\hat{a}_i^\dagger$ are bosonic annihilation and creation operators at lattice site $i$, $\hat{n}_i = \hat{a}_i^\dagger\hat{a}_i$ is the number operator, $J$ is the tunneling amplitude between nearest-neighbor sites $\langle i,j\rangle$, and $U$ is the on-site interaction strength.

To obtain the Anti-Wick symbol, we first rewrite the Hamiltonian in Anti-Wick order (all annihilation operators to the left of creation operators), using $[\hat{a}_i,\hat{a}_j^\dagger] = \delta_{ij}$.

We obtain:
\begin{equation}
\hat{H}_{\mathrm{BH}} = -J\sum_{\langle i,j\rangle}(\hat{a}_j\hat{a}_i^\dagger - \delta_{ij}) + \frac{U}{2}\sum_i(\hat{a}_i\hat{a}_i\hat{a}_i^\dagger\hat{a}_i^\dagger -3 \hat{a}_i\hat{a}_i^\dagger +2)\,.
\end{equation}

The Anti-Wick symbol is obtained by making the replacements $\hat{a}_i \to \alpha_i$ and $\hat{a}_i^\dagger \to \alpha_i^*$:
\begin{eqnarray}
&&H_{\mathrm{aW}}(\boldsymbol{\alpha},\boldsymbol{\alpha}^*) \\&=& -J\sum_{\langle i,j\rangle}\alpha_j\alpha_i^* \nonumber+ \frac{U}{2}\sum_i(\alpha_i^2\alpha_i^{*2} -3 \alpha_i^*\alpha_i)
+const.
\end{eqnarray}
The quartic interaction term $\alpha_i^2\alpha_i^{*2} = |\alpha_i|^4$ depends only on the local density $|\alpha_i|^2$ at each site. The quadratic term $-3\alpha_i^*\alpha_i = -3|\alpha_i|^2$ can be absorbed into a redefinition of the chemical potential. All third derivatives of $H_{\mathrm{aW}}$ vanish, the truncation condition Eq.~(\ref{condition}) is satisfied, and the Husimi evolution takes Fokker-Planck form. The framework developed in Sections~\ref{sec:derivation} and \ref{sec:Drummond} therefore applies directly to this experimentally realized interacting many-body system.

\subsection{Complex scalar field: the Higgs boson}\label{subsec:higgs}

The Higgs field in the Standard Model is a complex scalar field with quartic self-interaction. To see why it violates the truncation condition, we consider a simplified single-mode Hamiltonian (neglecting gauge couplings and spatial gradients):
\begin{equation}
\mathcal{H}[\phi,\pi]
=
\pi^* \pi
+ m|\phi|^2
+
\frac{\lambda}{2}\,|\phi|^4\,.
\end{equation}

For a complex field mode $\phi$, we introduce two independent complex phase-space variables: $\alpha$ with conjugate $\alpha^*$, and $\beta$ with conjugate $\beta^*$ (see \citep{drummondreid2021entropy}). We define
\begin{equation}
\phi = \frac{1}{\sqrt{2m}}\left(\alpha+\beta^*\right)\,,
\qquad
\pi = i\sqrt{\frac{m}{2}}\left(\beta-\alpha^*\right)\,.
\end{equation}
The quartic self-interaction term becomes
\begin{equation}
|\phi|^4 = \frac{1}{4m^2}(|\alpha|^2 + |\beta|^2 + \alpha\beta + \alpha^*\beta^*)^2\,,
\end{equation}
which has non-vanishing third derivatives with respect to the combined variables $\alpha$ and $\beta$ on the one hand and $\alpha^*$ and $\beta^*$ on the other. Unlike the Bose-Hubbard interaction, this is not a pure density-density coupling in the complex phase space variables. Correspondingly, the Husimi evolution includes terms beyond Fokker-Planck form.

\subsection{Gauge boson self-interactions}\label{subsec:gauge}

Non-Abelian gauge theories such as QCD feature cubic and quartic self-interactions of the gauge bosons. The QCD Lagrangian for gluon fields is
\begin{equation}
\mathcal{L}_{\mathrm{gluon}} = -\frac{1}{4}F^a_{\mu\nu}F^{a\mu\nu}\,,
\end{equation}
where the field strength tensor is
\begin{equation}
F^a_{\mu\nu} = \partial_\mu A^a_\nu - \partial_\nu A^a_\mu + g f^{abc} A^b_\mu A^c_\nu\,.
\end{equation}
Here $a,b,c = 1,\ldots,8$ are color indices for the adjoint representation of $\mathrm{SU}(3)$, and $f^{abc}$ are the structure constants of the Lie algebra.

Expanding the Lagrangian yields cubic and quartic terms in the gauge fields. After mode expansion 
\begin{equation}
A^a_\mu(\mathbf{x}) = \sum_k \bigl(\alpha_k^a u_{k\mu}(\mathbf{x}) + \alpha_k^{a*} u_{k\mu}^*(\mathbf{x})\bigr)\,,
\end{equation}
these become polynomials in the complex phase-space variables $\alpha_k^a$ and their conjugates. While the antisymmetry of the structure constants prevents any single mode amplitude $\alpha_k^a$ from appearing with power higher than two, there are mixed third and fourth derivatives with respect to different mode variables. The Husimi evolution therefore goes beyond Fokker-Planck form.

\subsection{Possible reactions to the limitation}\label{subsec:reactions}

The Fokker-Planck ansatz (constraint C2) was motivated by the expectation that underlying stochastic trajectories should be continuous. The fact that the Husimi evolution in some empirically important quantum field theories cannot be recovered in Fokker-Planck form raises the question of how to respond. We consider three possible reactions:

\subsubsection{Reaction 1: Go beyond Fokker-Planck form}

One may drop constraint C2 and allow terms in the evolution of the probability density beyond the Fokker-Planck stage. The identification of Fokker-Planck equations with continuous trajectories applies only to Markovian models. Since the correct underlying micro-dynamics of quantum field theories may be non-Markovian (as we substantiate in our companion paper), an account in terms of continuous micro-trajectories may be possible even when the probability density evolution is not confined to Fokker-Planck form. Alternatively, one might be willing to consider potentially discontinuous stochastic trajectories.

\subsubsection{Reaction 2: Adopt the framework of Gaussian bosonic operators}

A framework of Gaussian bosonic operators, which generalize coherent state projectors, has been developed by Corney and Drummond \citep{corneydrummond}. This framework features an enlarged phase space that allows one to describe the time evolution of the probability density for generic quartic interaction terms as a Fokker-Planck equation. Adopting this framework could enable the present approach to cover all bosonic fields in the Standard Model and has been suggested as a route to treating fermions as well \citep{drummondreid2021entropy}.

However, enlarging the phase space beyond the classical coordinates may break the natural connection between deterministic/free and stochastic/interacting dynamics established by constraint C7. Moreover, the appealing connection to quantization described in Section~\ref{sec:Drummond} might also not survive. It remains unclear whether an interpretation that relies on the Gaussian operator framework could have an equally compelling motivation from first principles as a generalization of classical mechanics.

\subsubsection{Reaction 3: Fokker-Planck truncation as an empirical alternative}

If the evolution equation Eq.~(\ref{eq:full_FPE_real}) admits a trajectory interpretation but the general Husimi evolution Eq.~(\ref{full_Q_evolution}) does not, one might consider theories in which the probability density evolution truncates at Fokker-Planck level -- even for Hamiltonians with non-vanishing higher derivatives -- as empirical alternatives to standard quantum field theory. Since beyond-Fokker-Planck terms represent higher-order quantum corrections, and since some Standard Model parameters (such as the Higgs self-coupling) have not yet been measured with high precision, such deviations might not have been empirically tested. This is highly speculative, but it illustrates how foundational investigations of trajectory interpretations may have empirical consequences and potentially make contact with future experiments.

\subsection{Recovering discrete measurement outcomes}\label{subsec:discrete_outcomes}

The preceding subsections addressed which quantum field theories fall within the scope of the Fokker-Planck framework. A more general concern about empirical adequacy arises even within this scope: can any framework that interprets the Husimi function as a proper probability density on phase space possibly account for the sharp, discrete measurement outcomes that are characteristic of quantum theory? After all, the Husimi function is a smooth, everywhere-positive probability density on phase space. How can 
such a distribution give rise to the appearance of definite eigenvalues upon measurement? We encountered a version of this concern in connection with constraint C7 (Section~\ref{subsec:scale_fixing}), which aligns the 
free/interacting and deterministic/stochastic distinctions: Quantum theory predicts discrete outcomes even for free systems, but the present framework says that free systems evolve deterministically, according to Hamilton's equations in phase space. How can these observations be reconciled?

The key insight is that measurement inevitably involves interaction 
processes. Whenever an observable is measured, the system of interest is coupled to a measurement apparatus through an amplification stage that supposedly magnifies microscopic differences into macroscopically distinguishable pointer positions. In our framework, this amplification interaction, which must in principle be modelled explicitly, generates stochastic dynamics and, thereby, inevitably involves non-classical elements.

Drummond and Reid \citep{drummondreid2020,drummondreid2026} have 
analysed this mechanism in detail for single-mode measurements. They 
consider a parametric amplification Hamiltonian of the form $H_{\mathrm{amp}} 
= \frac{i\hbar g}{2}(\hat{a}^{\dagger 2} - \hat{a}^2)$, which is quadratic in the mode operators and therefore falls within the scope of the present framework. They show that the amplified quadrature -- $q$ or $p$ or some linear combination -- evolves under the traceless diffusion dynamics in a bimodal way, with the two peaks corresponding to the two possible measurement outcomes, and their relative weights given by the Born rule applied to the quantum state at the end of the measurement process. The amplification process thus converts a unimodal Husimi function of the pre-measurement quantum state into a distribution with well-separated peaks -- recovering the appearance of discrete outcomes from continuous stochastic dynamics.

Complementarily, in a Hilbert-space analysis that does not rely on any specific microdynamics, one of us has shown \citep{friederich2024} 
that decoherence induced by the measurement apparatus leads to a 
Husimi function for the macroscopic ``pointer observable'' that is extremely well approximated by a mixture of Gaussians 
centred on the eigenvalue-associated coherent states, with weights 
given by the Born rule. These findings suggest that the $Q$-function framework can account for the empirical appearance of sharp measurement outcomes: what looks like a ``collapse'' to a sharp 
eigenvalue is the result of the measurement interaction concentrating the probability density of the macroscopic pointer or display configuration within specific phase-space regions. This suggests that the principles suggested here, which align the dichotomy free/interacting with the dichotomy deterministic/stochastic, can in principle account for sharp eigenvalues as measurement outcomes occurring with Born rule probabilities.

\section{Summary and outlook}\label{sec:summary}

We have derived a unique generalization of the Liouville equation of classical statistical mechanics by imposing seven physically motivated constraints: classical limit, Fokker-Planck form, local Hamiltonian dependence, time-reversal invariance, energy conservation, minimality, and alignment of the free/interacting and deterministic/stochastic distinctions. The resulting evolution equation for the probability density features a traceless diffusion matrix -- a structure that balances forward and backward diffusion to preserve time-reversal symmetry.

Remarkably, this equation coincides exactly with the Schr\"odinger equation in the coherent-state representation for bosonic quantum field theories whose Anti-Wick Hamiltonian symbols are at most quadratic in each phase-space variable. This correspondence fixes the otherwise free parameter $C$ in our derivation: $C = 1$. The other physical parameter $\kappa=\hbar$ is fixed by requiring that for non-interacting Hamiltonians the evolution of $\rho$ reduces to the classical Liouvillean evolution. The physical theories satisfying this condition include all free bosonic field theories and interacting theories with density-density couplings in the complex phase space variables, such as the experimentally realized Bose-Hubbard model.

This formal correspondence suggests interpreting the Husimi function as a genuine probability density on phase space rather than merely a quasiprobability distribution. Under such an interpretation, all dynamical variables would have sharp values at each instant, and quantum statistics would arise from incomplete knowledge rather than fundamental indefiniteness. However, establishing whether this probability density can be understood as arising from underlying stochastic trajectories requires analysis of the micro-dynamics -- a question we address in a companion paper.

\subsection{Limitations and open questions}

Several important issues remain:

\textbf{Scope of the framework.} Generic interaction terms in the Standard Model -- including the Higgs self-coupling and non-Abelian gauge boson interactions -- lead to Hamiltonians whose Anti-Wick symbols violate Eq.~(\ref{condition}). For such theories, the Husimi function time evolution includes derivatives beyond second order, taking it outside the Fokker-Planck framework developed here. Possible responses include: (1) abandoning constraint C2 and allowing higher-order terms in the probability evolution, (2) adopting the enlarged phase space of Gaussian bosonic operators \citep{corneydrummond}, or (3) investigating whether Fokker-Planck truncation provides empirically adequate alternatives to standard quantum field theory. Each option presents distinct theoretical and empirical challenges.

\textbf{Trajectory interpretation.} Whether the evolution equation derived here supports a consistent interpretation in terms of stochastic trajectories is, at this stage, an open question. Our companion paper investigates this question and investigates a proposal by Drummond \citep{drummond2021}. It is shown there the dynamics suggested by Drummond are fundamentally non-Markovian, which, encouragingly, places them outside the scope of standard no-go theorems for non-locality, contextuality, and the reality of the wave function $\psi$. We also point out significant interpretive challenges that remain for Drummond's proposal.

\textbf{Extension to fermions.} The present analysis covers only bosonic fields. While Drummond \citep{drummond2021} has argued that traceless diffusion Fokker-Planck equations also govern fermionic Q-function evolution \citep{rosales2015}, integrating fermions into the present framework remains an open problem.

\textbf{Relativistic generalization.} The traceless diffusion Fokker-Planck equation involves a preferred time coordinate, raising questions about Lorentz covariance. Whether a manifestly covariant formulation exists, or whether foliation choices should be understood as physical (as in Bohmian mechanics), requires further investigation.

\subsection{Significance}

The central achievement of this work is demonstrating that time-reversal invariance, energy conservation, and other natural physical constraints uniquely determine a stochastic generalization of classical mechanics -- and that this generalization reproduces quantum field theory for a significant class of bosonic systems. This may be seen as providing a certain amount of renewed support for Einstein's vision of quantum theory as fundamentally analogous to classical statistical mechanics, with probability distributions representing incomplete knowledge rather than ontological indeterminacy. Whether this vision can be fully realized -- encompassing all empirically confirmed quantum phenomena including fermions, relativistic effects, and the complete Standard Model -- remains an open question. In Part II of the paper we embark on the project of providing an interpretation of the equation derived here in terms of stochastic trajectories.

\section*{Acknowledgments}
We would like to thank Benjamin Feintzeig for helpful feedback of an earlier version. We initially used OpenAI's GPT 5 for obtaining feedback on our ideas. Later, we used Anthropic's Claude Opus 4.6 for assistance with formulations.

This research was funded by the Netherlands Organization for Scientific Research (NWO), project VI.Vidi.211.088.

\appendix

\section{Transformation from complex to real phase-space coordinates}
\label{app:complex_to_real}

We now derive the commutator expression $D(z) = \frac{C}{2\kappa^2}[\tilde{H}''(z), J]$ from the complex-variable result $D_{ij}^{\alpha\alpha} = C\frac{\partial^2 H}{\partial \alpha_i^* \partial \alpha_j^*}$.

From Eq.\ (\ref{complex_to_real}), the complex variables $\alpha_j = \sqrt\frac{1\kappa}{2}(Q_j + iP_j)$ satisfy
\begin{align}
\frac{\partial}{\partial \alpha_j} &= \frac{1}{\sqrt{2\kappa}}\left(\frac{\partial}{\partial Q_j} - i\frac{\partial}{\partial P_j}\right)\!,\label{dalpha}\\
\frac{\partial}{\partial \alpha_j^*} &= \frac{1}{\sqrt{2\kappa}}\left(\frac{\partial}{\partial Q_j} + i\frac{\partial}{\partial P_j}\right)\!.\label{dalphastar}
\end{align}

The diffusion matrix in complex variables has the structure
\begin{equation}
D^{\alpha\alpha}_{ij} = C\frac{\partial^2 H}{\partial \alpha_i^* \partial \alpha_j^*}\,.
\end{equation}

Applying the chain rule using Eq.\ (\ref{dalphastar}):
\begin{equation}
\resizebox{\columnwidth}{!}{%
    $\displaystyle
\begin{aligned}
\frac{\partial^2 H}{\partial \alpha_i^* \partial \alpha_j^*} 
&= \frac{1}{2\kappa}\left(\frac{\partial}{\partial Q_i} + i\frac{\partial}{\partial P_i}\right)\left(\frac{\partial}{\partial Q_j} + i\frac{\partial}{\partial P_j}\right)H\nonumber\\
&= \frac{1}{2\kappa}\left(\frac{\partial^2 H}{\partial Q_i \partial Q_j} + i\frac{\partial^2 H}{\partial Q_i \partial P_j} + i\frac{\partial^2 H}{\partial P_i \partial Q_j} - \frac{\partial^2 H}{\partial P_i \partial P_j}\right)\!.\label{second_deriv_complex}
\end{aligned}
$}
\end{equation}
In matrix notation, let us denote the Hessian in dimensionless real variables as
\begin{equation}
\tilde{H}''(z) = \begin{pmatrix}
\tilde{H}''_{QQ} & \tilde{H}''_{QP}\\
\tilde{H}''_{PQ} & \tilde{H}''_{PP}
\end{pmatrix}\!,
\end{equation}
where each block is an $N \times N$ matrix with entries
\begin{align}
(\tilde{H}''_{QQ})_{ij} &= \frac{\partial^2 H}{\partial Q_i \partial Q_j}\,,\quad
(\tilde{H}''_{QP})_{ij} = \frac{\partial^2 H}{\partial Q_i \partial P_j}\,,\\
(\tilde{H}''_{PQ})_{ij} &= \frac{\partial^2 H}{\partial P_i \partial Q_j}\,,\quad
(\tilde{H}''_{PP})_{ij} = \frac{\partial^2 H}{\partial P_i \partial P_j}\,.
\end{align}

Using Eq.\ (\ref{second_deriv_complex}) we write
\begin{equation}
\resizebox{\columnwidth}{!}{%
    $\displaystyle
D^{\alpha\alpha}_{ij} = \frac{C}{2\kappa}\left(\tilde{H}''_{QQ} - \tilde{H}''_{PP} + i(\tilde{H}''_{QP} + \tilde{H}''_{PQ})\right)_{ij}\!.\label{D_alpha_alpha_blocks}
$}
\end{equation}

We now derive the diffusion matrix in real $(Q,P)$ coordinates:
\begin{equation}
D(z) = \begin{pmatrix}
D_{QQ} & D_{QP}\\
D_{PQ} & D_{PP}
\end{pmatrix}\!.
\end{equation}

The diffusion contribution to the Fokker-Planck equation, Eq.~(\ref{diffusion_simplified}), with $D^{\alpha\alpha^*}=0$ reads
\begin{equation}
\resizebox{\columnwidth}{!}{%
    $\displaystyle
\mathcal{D}\rho = \frac{i\hbar}{2}\sum_{i,j}\left\{\frac{\partial^2}{\partial\alpha_i\partial\alpha_j}(D^{\alpha\alpha}_{ij}\rho) - \frac{\partial^2}{\partial\alpha_i^*\partial\alpha_j^*}((D^{\alpha\alpha}_{ij})^*\rho)\right\}\!.
$}
\end{equation}

To convert to real variables, we substitute the chain rule expressions Eqs.~(\ref{dalpha})--(\ref{dalphastar}) and $D^{\alpha\alpha} = \frac{C}{2\kappa}(A + iB)$, where
\begin{equation}
A_{ij} = (\tilde{H}''_{QQ} - \tilde{H}''_{PP})_{ij}\,,\quad B_{ij} = (\tilde{H}''_{QP} + \tilde{H}''_{PQ})_{ij}\,.
\end{equation}
Note that for a real-valued Hamiltonian, both $A$ and $B$ are real symmetric matrices, and the symmetry of the Hessian gives $\tilde{H}''_{QP} = (\tilde{H}''_{PQ})^{\mathsf{T}}$.

Expanding the second-order differential operators:
\begin{align}
\frac{\partial^2}{\partial\alpha_i\partial\alpha_j} &= \frac{1}{2\kappa}\bigl(\partial_{Q_i}\partial_{Q_j} - i\partial_{Q_i}\partial_{P_j} - i\partial_{P_i}\partial_{Q_j} - \partial_{P_i}\partial_{P_j}\bigr)\!,\nonumber\\
\frac{\partial^2}{\partial\alpha_i^*\partial\alpha_j^*} &= \frac{1}{2\kappa}\bigl(\partial_{Q_i}\partial_{Q_j} + i\partial_{Q_i}\partial_{P_j} + i\partial_{P_i}\partial_{Q_j} - \partial_{P_i}\partial_{P_j}\bigr)\!.\nonumber
\end{align}

The diffusion operator becomes
\begin{widetext}
\begin{align}
\mathcal{D}\rho &= \frac{i\hbar}{2}\cdot\frac{C}{4\kappa^2}\sum_{i,j}\Bigl\{
(\partial_{Q_i}\partial_{Q_j} - i\partial_{Q_i}\partial_{P_j} - i\partial_{P_i}\partial_{Q_j} - \partial_{P_i}\partial_{P_j})\bigl[(A_{ij}+iB_{ij})\rho\bigr]\nonumber\\
&\qquad\qquad\qquad - (\partial_{Q_i}\partial_{Q_j} + i\partial_{Q_i}\partial_{P_j} + i\partial_{P_i}\partial_{Q_j} - \partial_{P_i}\partial_{P_j})\bigl[(A_{ij}-iB_{ij})\rho\bigr]
\Bigr\}\!.
\end{align}
\end{widetext}
Taking the difference yields:
\begin{align}
\mathcal{D}\rho &= \frac{i\hbar C}{4\kappa^2}\sum_{i,j}\Bigl\{
i(\partial_{Q_i}\partial_{Q_j} - \partial_{P_i}\partial_{P_j})(B_{ij}\rho)\nonumber\\
&\qquad - i(\partial_{Q_i}\partial_{P_j} + \partial_{P_i}\partial_{Q_j})(A_{ij}\rho)
\Bigr\}\nonumber\\
&= -\frac{\hbar C}{4\kappa^2}\sum_{i,j}\Bigl\{
(\partial_{Q_i}\partial_{Q_j} - \partial_{P_i}\partial_{P_j})(B_{ij}\rho)\nonumber\\
&\qquad - (\partial_{Q_i}\partial_{P_j} + \partial_{P_i}\partial_{Q_j})(A_{ij}\rho)
\Bigr\}\!.\label{diffusion_real_expanded}
\end{align}
Using the definitions $A_{ij} = (\tilde{H}''_{QQ} - \tilde{H}''_{PP})_{ij}$ and $B_{ij} = (\tilde{H}''_{QP}+\tilde {H}''_{PQ})_{ij}$ and noting that for a real Hamiltonian the full Hessian is symmetric, so $\tilde{H}''_{PQ} = (\tilde{H}''_{QP})^T$ gives

\begin{align}
\mathcal{D}\rho &= -\frac{\hbar C}{4\kappa^2}\sum_{i,j}\Bigl\{
\partial_{Q_i}\partial_{Q_j}\bigl((\tilde{H}''_{QP}+\tilde{H}''_{PQ})_{ij}\rho\bigr)\nonumber\\
&\quad - (\partial_{Q_i}\partial_{P_j} + \partial_{P_i}\partial_{Q_j})\bigl((\tilde{H}''_{QQ} - \tilde{H}''_{PP})_{ij}\rho\bigr)\nonumber\\
&\quad - \partial_{P_i}\partial_{P_j}\bigl((\tilde{H}''_{QP}+\tilde{H}''_{PQ})_{ij}\rho\bigr)
\Bigr\}\!.
\end{align}

This has the Fokker-Planck form $\mathcal{D}\rho = \frac{\hbar}{2}\sum_{a,b}\tilde{\partial}_a\tilde{\partial}_b(D_{ab}\rho)$
, from which we identify:

\begin{align}
D_{QQ} &= -\frac{C}{2\kappa^2}\bigl(\tilde{H}''_{QP}+\tilde{H}''_{PQ}\bigr)\,,\label{DQQ}\\
D_{QP} = D_{PQ} &= \frac{C}{2\kappa^2}\bigl(\tilde{H}''_{QQ} - \tilde{H}''_{PP}\bigr)\,,\label{DQP}\\
D_{PP} &= \frac{C}{2\kappa^2}\bigl(\tilde{H}''_{QP}+\tilde{H}''_{PQ}\bigr)\,.\label{DPP}
\end{align}
Note that $D_{QP} = D_{PQ}$, confirming the symmetry of the diffusion matrix as required by constraint C2.

We now verify that this equals $\frac{C}{2\kappa^2}[\tilde{H}'', J]$. Computing the commutator:
\begin{equation}
\resizebox{\columnwidth}{!}{%
    $\displaystyle
\begin{aligned}
[\tilde{H}'', J] &= \begin{pmatrix}
\tilde{H}''_{QQ} & \tilde{H}''_{QP}\\
\tilde{H}''_{PQ} & \tilde{H}''_{PP}
\end{pmatrix}
\begin{pmatrix}
0 & I_N\\ -I_N & 0
\end{pmatrix}
- \begin{pmatrix}
0 & I_N\\ -I_N & 0
\end{pmatrix}
\begin{pmatrix}
\tilde{H}''_{QQ} & \tilde{H}''_{QP}\\
\tilde{H}''_{PQ} & \tilde{H}''_{PP}
\end{pmatrix}\nonumber\\
&= \begin{pmatrix}
-\tilde{H}''_{QP} & \tilde{H}''_{QQ}\\
-\tilde{H}''_{PP} & \tilde{H}''_{PQ}
\end{pmatrix}
- \begin{pmatrix}
\tilde{H}''_{PQ} & \tilde{H}''_{PP}\\
-\tilde{H}''_{QQ} & -\tilde{H}''_{QP}
\end{pmatrix}\nonumber\\
&= \begin{pmatrix}
-\tilde{H}''_{QP} - \tilde{H}''_{PQ} & \tilde{H}''_{QQ} - \tilde{H}''_{PP}\\
\tilde{H}''_{QQ} - \tilde{H}''_{PP} & \tilde{H}''_{PQ} + \tilde{H}''_{QP}
\end{pmatrix}\!.\label{commutator_explicit}
\end{aligned}
$}
\end{equation}

This matrix is manifestly symmetric, with equal off-diagonal blocks $\tilde{H}''_{QQ} - \tilde{H}''_{PP}$. Comparing with Eqs.~(\ref{DQQ})--(\ref{DPP}), we confirm
\begin{equation}
D(z) = \frac{C}{2\kappa^2}[\tilde{H}''(z), J]\,,
\end{equation}
as claimed.

\bibliographystyle{apsrev4-2} 
\bibliography{bibliography}

@misc{appleby1998optimaljointmeasurementsposition,
      title={Optimal Joint Measurements of Position and Momentum}, 
      author={D. M. Appleby},
      year={1998},
      eprint={quant-ph/9803053},
      archivePrefix={arXiv},
      primaryClass={quant-ph},
      url={https://arxiv.org/abs/quant-ph/9803053}, 
}

@article{Appleby_2000,
doi = {10.1088/0305-4470/33/21/304},
url = {https://dx.doi.org/10.1088/0305-4470/33/21/304},
year = {2000},
month = {jun},
publisher = {},
volume = {33},
number = {21},
pages = {3903},
author = {D M Appleby},
title = {Husimi transform of an
operator product},
journal = {Journal of Physics A: Mathematical and General}
}

@incollection{bacciagaluppi2012,
  author = {Bacciagaluppi, Guido},
  title = {A conceptual introduction to {N}elson's mechanics},
  booktitle = {Endophysics, Time, Quantum and the Subjective},
  editor = {Buccheri, R. and Saniga, M. and Elitzur, A.},
  publisher = {World Scientific},
  address = {Singapore},
  pages = {367--388},
  year = {2005},
  note = {Revised version available at \url{https://philsci-archive.pitt.edu/8853/}}
}

@article{bell1964,
  author = {Bell, John S.},
  title = {On the Einstein-Podolsky-Rosen paradox},
  journal = {Physics},
  volume = {1},
  pages = {195--200},
  year = {1964}
}

@article{bopp1956,
  author = {Bopp, Fritz},
  title = {La m{\'e}canique quantique est-elle une m{\'e}canique statistique classique particuli{\`e}re?},
  journal = {Annales de l'Institut Henri Poincar{\'e}},
  volume = {15},
  pages = {81--112},
  year = {1956}
}

@article{corneydrummond,
  title = {Gaussian quantum operator representation for bosons},
  author = {Corney, Joel F. and Drummond, Peter D.},
  journal = {Phys. Rev. A},
  volume = {68},
  issue = {6},
  pages = {063822},
  numpages = {22},
  year = {2003},
  month = {Dec},
  publisher = {American Physical Society},
  doi = {10.1103/PhysRevA.68.063822},
  url = {https://link.aps.org/doi/10.1103/PhysRevA.68.063822}
}

@article{daubechiesklauder1985,
  author = {Daubechies, Ingrid and Klauder, John R.},
  title = {Quantum-mechanical path integrals with {W}iener measures for all polynomial {H}amiltonians. {II}},
  journal = {Journal of Mathematical Physics},
  volume = {26},
  pages = {2239--2256},
  year = {1985},
  doi = {10.1063/1.526803}
}

@article{drummond2021,
  author = {Drummond, Peter D.},
  title = {Time evolution with symmetric stochastic action},
  journal = {Physical Review Research},
  volume = {3},
  pages = {013240},
  year = {2021},
  doi = {10.1103/PhysRevResearch.3.013240}
}

@article{drummondreid2020,
  author = {Drummond, Peter D. and Reid, Margaret D.},
  title = {Retrocausal model of reality for quantum fields},
  journal = {Physical Review Research},
  volume = {2},
  pages = {033266},
  year = {2020},
  doi = {10.1103/PhysRevResearch.2.033266}
}

@article{drummondreid2021entropy,
  author = {Drummond, Peter D. and Reid, Margaret D.},
  title = {Objective Quantum Fields, Retrocausality and Ontology},
  journal = {Entropy},
  volume = {23},
  number = {6},
  pages = {749},
  year = {2021},
  doi = {10.3390/e23060749}
}

@article{drummondreid2026,
  author = {Reid, Margaret D. and Drummond, Peter D.},
  title = {Forward-backward stochastic simulations: {Q}-based model for measurement and {B}ell nonlocality consistent with weak local realistic premises},
  journal = {Physical Review A},
  volume = {113},
  pages = {012210},
  year = {2026},
  doi = {10.1103/PhysRevA.113.012210}
}

@incollection{einstein1949,
  author = {Einstein, Albert},
  title = {Remarks concerning the essays brought together in this cooperative volume},
  booktitle = {Albert Einstein: Philosopher-Scientist},
  editor = {Schilpp, Paul Arthur},
  series = {The Library of Living Philosophers},
  volume = {7},
  pages = {665--688},
  publisher = {Library of Living Philosophers},
  address = {Evanston, IL},
  year = {1949}
}

@article{fenyes1952,
  author = {F{\'e}nyes, Imre},
  title = {Eine wahrscheinlichkeitstheoretische {B}egr{\"u}ndung und {I}nterpretation der {Q}uantenmechanik},
  journal = {Zeitschrift f{\"u}r Physik},
  volume = {132},
  pages = {81--106},
  year = {1952},
  doi = {10.1007/BF01338578}
}

@book{folland1989,
  author = {Folland, Gerald B.},
  title = {Harmonic Analysis in Phase Space},
  publisher = {Princeton University Press},
  address = {Princeton},
  year = {1989}
}

@article{friederich2024,
  author = {Friederich, Simon},
  title = {Introducing the {Q}-based interpretation of quantum mechanics},
  journal = {British Journal for the Philosophy of Science},
  volume = {75},
  pages = {769--795},
  year = {2024},
  doi = {10.1086/714810}
}

@article{depolavieja1996,
title = {A causal quantum theory in phase space},
journal = {Physics Letters A},
volume = {220},
number = {6},
pages = {303-314},
year = {1996},
issn = {0375-9601},
doi = {https://doi.org/10.1016/0375-9601(96)00523-3},
url = {https://www.sciencedirect.com/science/article/pii/0375960196005233},
author = {Gonzalo {García de Polavieja}}
}

@article{goldstein2011,
  author = {Goldstein, Sheldon and Norsen, Travis and Tausk, Daniel Victor and Zangh{\`\i}, Nino},
  title = {Bell's theorem},
  journal = {Scholarpedia},
  volume = {6},
  number = {10},
  pages = {8378},
  year = {2011},
  doi = {10.4249/scholarpedia.8378}
}

@article{grabert1979,
  author = {Grabert, Hermann and H{\"a}nggi, Peter and Talkner, Peter},
  title = {Is quantum mechanics equivalent to a classical stochastic process?},
  journal = {Physical Review A},
  volume = {19},
  pages = {2440--2445},
  year = {1979},
  doi = {10.1103/PhysRevA.19.2440}
}

@book{hall2013,
  author = {Hall, Brian C.},
  title = {Quantum Theory for Mathematicians},
  series = {Graduate Texts in Mathematics},
  volume = {267},
  publisher = {Springer},
  address = {0},
  year = {2013},
  doi = {},
  isbn = {}
}

@incollection{held2022,
  author = {Held, Carsten},
  title = {The {K}ochen-{S}pecker Theorem},
  booktitle = {The Stanford Encyclopedia of Philosophy},
  editor = {Zalta, Edward N. and Nodelman, Uri},
  edition = {Fall 2022},
  year = {2022},
  publisher = {Metaphysics Research Lab, Stanford University},
  url = {https://plato.stanford.edu/archives/fall2022/entries/kochen-specker/}
}

@article{husimi1940,
  author = {Husimi, K{\^o}di},
  title = {Some formal properties of the density matrix},
  journal = {Proceedings of the Physico-Mathematical Society of Japan},
  volume = {22},
  pages = {264--314},
  year = {1940}
}

@article{kochen1967,
  author = {Kochen, Simon and Specker, Ernst P.},
  title = {The problem of hidden variables in quantum mechanics},
  journal = {Journal of Mathematics and Mechanics},
  volume = {17},
  pages = {59--87},
  year = {1967}
}

@article{lee,
title = {Theory and application of the quantum phase-space distribution functions},
journal = {Physics Reports},
volume = {259},
number = {3},
pages = {147-211},
year = {1995},
issn = {0370-1573},
doi = {https://doi.org/10.1016/0370-1573(95)00007-4},
url = {https://www.sciencedirect.com/science/article/pii/0370157395000074},
author = {Hai-Woong Lee}
}

@article{leifer2014,
  author = {Leifer, Matthew S.},
  title = {Is the quantum state real? {A}n extended review of $\psi$-ontology theorems},
  journal = {Quanta},
  volume = {3},
  pages = {67--155},
  year = {2014},
  doi = {10.12743/quanta.v3i1.22}
}

@article{miranker1961,
  author = {Miranker, Willard L.},
  title = {A well posed problem for the backward heat equation},
  journal = {Proceedings of the American Mathematical Society},
  volume = {12},
  pages = {243--247},
  year = {1961},
  doi = {10.1090/S0002-9939-1961-0120462-2}
}

@article{MIZRAHI1984241,
title = {Quantum mechanics in the Gaussian wave-packet phase space representation},
journal = {Physica A: Statistical Mechanics and its Applications},
volume = {127},
number = {1},
pages = {241-264},
year = {1984},
issn = {0378-4371},
doi = {https://doi.org/10.1016/0378-4371(84)90130-4},
url = {https://www.sciencedirect.com/science/article/pii/0378437184901304},
author = {S.S. Mizrahi}
}

@article{MIZRAHI1986237,
title = {Quantum mechanics in the Gaussian wave-packet phase space representation II: Dynamics},
journal = {Physica A: Statistical Mechanics and its Applications},
volume = {135},
number = {1},
pages = {237-250},
year = {1986},
issn = {0378-4371},
doi = {https://doi.org/10.1016/0378-4371(86)90115-9},
url = {https://www.sciencedirect.com/science/article/pii/0378437186901159},
author = {S.S. Mizrahi}
}

@article{PhysRevLett.117.070801,
  title = {Evading Vacuum Noise: Wigner Projections or Husimi Samples?},
  author = {M\"uller, C. R. and Peuntinger, C. and Dirmeier, T. and Khan, I. and Vogl, U. and Marquardt, Ch. and Leuchs, G. and S\'anchez-Soto, L. L. and Teo, Y. S. and Hradil, Z. and \ifmmode \check{R}\else \v{R}\fi{}eh\'a\ifmmode \check{c}\else \v{c}\fi{}ek, J.},
  journal = {Phys. Rev. Lett.},
  volume = {117},
  issue = {7},
  pages = {070801},
  numpages = {6},
  year = {2016},
  month = {Aug},
  publisher = {American Physical Society},
  doi = {10.1103/PhysRevLett.117.070801},
  url = {https://link.aps.org/doi/10.1103/PhysRevLett.117.070801}
}

@incollection{myrvold2024,
  author = {Myrvold, Wayne and Genovese, Marco and Shimony, Abner},
  title = {Bell's theorem},
  booktitle = {The Stanford Encyclopedia of Philosophy},
  editor = {Zalta, Edward N. and Nodelman, Uri},
  edition = {Spring 2024},
  year = {2024},
  publisher = {Metaphysics Research Lab, Stanford University},
  url = {https://plato.stanford.edu/archives/spr2024/entries/bell-theorem/}
}

@article{nelson1966,
  author = {Nelson, Edward},
  title = {Derivation of the {S}chr{\"o}dinger equation from {N}ewtonian mechanics},
  journal = {Physical Review},
  volume = {150},
  pages = {1079--1085},
  year = {1966},
  doi = {10.1103/PhysRev.150.1079}
}

@article{pusey2012,
  author = {Pusey, Matthew F. and Barrett, Jonathan and Rudolph, Terry},
  title = {On the reality of the quantum state},
  journal = {Nature Physics},
  volume = {8},
  pages = {475--478},
  year = {2012},
  doi = {10.1038/nphys2309}
}

@article{rosales2015,
  author = {Rosales-Z{\'a}rate, Laura E. C. and Drummond, Peter D.},
  title = {Probabilistic {Q}-function distributions in fermionic phase-space},
  journal = {New Journal of Physics},
  volume = {17},
  pages = {032002},
  year = {2015},
  doi = {10.1088/1367-2630/17/3/032002}
}

@article{torresvegafrederick,
    author = {Torres‐Vega, Go. and Frederick, John H.},
    title = {Quantum mechanics in phase space: New approaches to the correspondence principle},
    journal = {The Journal of Chemical Physics},
    volume = {93},
    number = {12},
    pages = {8862-8874},
    year = {1990},
    month = {12},

   
    issn = {0021-9606},
    doi = {10.1063/1.459225},
    url = {https://doi.org/10.1063/1.459225},
    eprint = {https://pubs.aip.org/aip/jcp/article-pdf/93/12/8862/18990687/8862\_1\_online.pdf},
}

@article{tyagifriederich,
  author = {Tyagi, Mritunjay and Friederich, Simon},
  title = {Time evolution of the {H}usimi and {G}lauber-{S}udarshan functions in terms of complementary {H}amiltonian symbols},
  journal = {arXiv preprint arXiv:2510.15628},
  year = {2025},
  note = {Preprint},
  eprint = {},
  archivePrefix = {},
  primaryClass = {quant-ph}
}

@article{wallstrom1994,
  author = {Wallstrom, Timothy C.},
  title = {Inequivalence between the {S}chr{\"o}dinger equation and the {M}adelung hydrodynamic equations},
  journal = {Physical Review A},
  volume = {49},
  pages = {1613--1617},
  year = {1994},
  doi = {10.1103/PhysRevA.49.1613}
}

@article{watanabe1965,
  author = {Watanabe, Satosi},
  title = {Conditional probability in physics},
  journal = {Progress of Theoretical Physics Supplement},
  volume = {E65},
  pages = {135--160},
  year = {1965},
  note = {Supplement dedicated to Yuzuru Watanabe}
}

@book{Wiseman_Milburn_2009, place={Cambridge}, title={Quantum Measurement and Control}, publisher={Cambridge University Press}, author={Wiseman, Howard M. and Milburn, Gerard J.}, year={2009}}

@article{fisher1989,
  title = {Boson localization and the superfluid-insulator transition},
  author = {Fisher, Matthew P. A. and Weichman, Peter B. and Grinstein, G. and Fisher, Daniel S.},
  journal = {Physical Review B},
  volume = {40},
  issue = {1},
  pages = {546--570},
  year = {1989},
  month = {Jul},
  publisher = {American Physical Society},
  doi = {10.1103/PhysRevB.40.546}
}

@article{jaksch1998,
  title = {Cold Bosonic Atoms in Optical Lattices},
  author = {Jaksch, D. and Bruder, C. and Cirac, J. I. and Gardiner, C. W. and Zoller, P.},
  journal = {Physical Review Letters},
  volume = {81},
  issue = {15},
  pages = {3108--3111},
  year = {1998},
  month = {Oct},
  publisher = {American Physical Society},
  doi = {10.1103/PhysRevLett.81.3108}
}

@article{greiner2002,
  title = {Quantum phase transition from a superfluid to a Mott insulator in a gas of ultracold atoms},
  author = {Greiner, Markus and Mandel, Olaf and Esslinger, Tilman and H\"ansch, Theodor W. and Bloch, Immanuel},
  journal = {Nature},
  volume = {415},
  pages = {39--44},
  year = {2002},
  month = {Jan},
  doi = {10.1038/415039a}
}

@article{bloch2008,
  title = {Many-body physics with ultracold gases},
  author = {Bloch, Immanuel and Dalibard, Jean and Zwerger, Wilhelm},
  journal = {Reviews of Modern Physics},
  volume = {80},
  issue = {3},
  pages = {885--964},
  year = {2008},
  month = {Jul},
  publisher = {American Physical Society},
  doi = {10.1103/RevModPhys.80.885}
}

@book{baezsegalzhou,
  author    = {J. C. Baez and I. E. Segal and Z. Zhhou},
  title     = {Introduction to Algebraic and Constructive Quantum Field Theory},
  publisher = {Princeton University Press},
  address   = {},
  year      = {1992}
}

@article{browning2020,
    author = {Browning, Thomas L. and Feintzeig, Benjamin H. and Gates-Redburg, Robin and Librande, Jonah and Soiffer, Rory},
    title = {Classical limits of unbounded quantities by strict quantization},
    journal = {Journal of Mathematical Physics},
    volume = {61},
    number = {11},
    pages = {112305},
    year = {2020},
    month = {11},
    issn = {0022-2488},
    doi = {10.1063/1.5142182},
    url = {https://doi.org/10.1063/1.5142182}
}

\end{document}